\documentclass[twocolumn,tighten,times,twocolappendix]{aastex631}

\begin{document}
\newcommand{\oi}{\text{[\ion{O}{1}]}}
\newcommand{\oii}{\text{[\ion{O}{2}]}}
\newcommand{\neiii}{\text{[\ion{Ne}{3}]}}
\newcommand{\oiii}{\text{[\ion{O}{3}]}}
\newcommand{\woiii}{\text{$W_\lambda(\oiii)$}}
\newcommand{\nii}{\text{[\ion{N}{2}]}}
\newcommand{\hei}{\text{\ion{He}{1}}}
\newcommand{\heii}{\text{\ion{He}{2}}}
\newcommand{\ha}{\text{H$\alpha$}}
\newcommand{\wha}{\text{$W_\lambda(\ha)$}}
\newcommand{\hb}{\text{H$\beta$}}
\newcommand{\hg}{\text{H$\gamma$}}
\newcommand{\hd}{\text{H$\delta$}}
\newcommand{\he}{\text{H$\epsilon$}}
\newcommand{\hz}{\text{H$\zeta$}}
\newcommand{\hn}{\text{H$\eta$}}
\newcommand{\htheta}{\text{H$\theta$}}
\newcommand{\hiota}{\text{H$\iota$}}
\newcommand{\pa}{\text{Pa$\alpha$}}
\newcommand{\pb}{\text{Pa$\beta$}}
\newcommand{\pg}{\text{Pa$\gamma$}}
\newcommand{\pd}{\text{Pa$\delta$}}
\newcommand{\hi}{\text{\ion{H}{1}}}
\newcommand{\hii}{\text{\ion{H}{2}}}
\newcommand{\hk}{\text{H$\kappa$}}
\newcommand{\caii}{\text{\ion{Ca}{2}}}
\newcommand{\sii}{\text{[\ion{S}{2}]}}
\newcommand{\siii}{\text{[\ion{S}{3}]}}
\newcommand{\wlya}{\text{$W_\lambda$({\rm Ly$\alpha$})}}
\newcommand{\wlyaem}{\text{$W_\lambda^{\rm em}$({\rm Ly$\alpha$})}}
\newcommand{\llya}{\text{$L$(Ly$\alpha$)}}
\newcommand{\llyaobs}{\text{$L$(Ly$\alpha$)$_{\rm obs}$}}
\newcommand{\llyaint}{\text{$L$(Ly$\alpha$)$_{\rm int}$}}
\newcommand{\lyafrac}{\text{$f_{\rm esc}^{\rm spec}$(Ly$\alpha$)}}
\newcommand{\lha}{\text{$L$(H$\alpha$)}}
\newcommand{\lhb}{\text{$L$(H$\beta$)}}
\newcommand{\sfrha}{\text{SFR(\ha)}}
\newcommand{\sfrneb}{{\rm SFR}_{\rm neb}}
\newcommand{\sfrsed}{\text{SFR(SED)}}
\newcommand{\sfruv}{\text{SFR(UV)}}
\newcommand{\ssfrha}{\text{sSFR(\ha)}}
\newcommand{\ssfrsed}{\text{sSFR(SED)}}
\newcommand{\ebmv}{E(B-V)}
\newcommand{\ebmvneb}{E(B-V)_{\rm neb}}
\newcommand{\ebmvcont}{E(B-V)_{\rm cont}}
\newcommand{\ebmvlos}{E(B-V)_{\rm los}}
\newcommand{\nhi}{N(\text{\ion{H}{1}})}
\newcommand{\lognhi}{\log[\nhi/{\rm cm}^{-2}]}
\newcommand{\lognhitable}{\log\left[\frac{\nhi}{{\rm cm}^{-2}}\right]}
\newcommand{\lya}{\text{Ly$\alpha$}}
\newcommand{\lyb}{\text{Ly$\beta$}}
\newcommand{\lyg}{\text{Ly$\gamma$}}
\newcommand{\comment}[1]{}
\newcommand{\wciii}{\text{$W_\lambda$(\ion{C}{3}])}}
\newcommand{\ciii}{\text{\ion{C}{3}]}}
\newcommand{\interoiii}{\text{\ion{O}{3}]}}
\newcommand{\rsiione}{R(\text{\ion{Si}{2}}\lambda 1260)}
\newcommand{\rsiitwo}{R(\text{\ion{Si}{2}}\lambda 1527)}
\newcommand{\rsii}{R(\text{\ion{S}{2}})}
\newcommand{\cii}{\text{\ion{C}{2}}}
\newcommand{\civ}{\text{\ion{C}{4}}}
\newcommand{\siiv}{\text{\ion{Si}{4}}}
\newcommand{\rcii}{R(\text{\ion{C}{2}}\lambda 1334)}
\newcommand{\ralii}{R(\ion{Al}{2}\lambda 1670)}
\newcommand{\qh}{Q(\text{H$^0$})}
\newcommand{\rs}{{\cal R}_{\rm s}}
\newcommand{\fcov}{f_{\rm cov}}
\newcommand{\fcovhi}{f_{\rm cov}(\hi)}
\newcommand{\fcovmetal}{f_{\rm cov}({\rm metal})}
\newcommand{\fesclya}{f_{\rm esc}^{\rm spec}(\lya)}
\newcommand{\logxi}{\log[\xi_{\rm ion}/{\rm s^{-1}/erg\,s^{-1}\,Hz^{-1}}]}
\newcommand{\logq}{\log[Q/{\rm s^{-1}}]}
\newcommand{\lir}{L_{\rm IR}}
\newcommand{\lbol}{L_{\rm bol}}
\newcommand{\luv}{L({\rm UV})}
\newcommand{\rv}{R_V}

\title{The JWST/AURORA Survey: Multiple Balmer and Paschen Emission Lines for Individual Star-forming Galaxies at $z=1.5-4.4$. I. A Diversity of Nebular Attenuation Curves and Evidence for Non-Unity Dust Covering Fractions}


\author[0000-0001-9687-4973]{Naveen A. Reddy}
\affiliation{Department of Physics and Astronomy, University of California, 
Riverside, 900 University Avenue, Riverside, CA 92521, USA; naveenr@ucr.edu}

\author[0000-0003-3509-4855]{Alice E. Shapley}
\affiliation{Department of Physics \& Astronomy, University of California,
Los Angeles, 430 Portola Plaza, Los Angeles, CA 90095, USA}

\author[0000-0003-4792-9119]{Ryan L. Sanders}\affiliation{Department of Physics and Astronomy, University of Kentucky, 505 Rose Street, Lexington, KY 40506, USA}

\author[0000-0001-8426-1141]{Michael W. Topping}
\affiliation{Steward Observatory, University of Arizona, 933 North 
Cherry Avenue, Tucson, AZ 85721, USA}

\author[0000-0001-7782-7071]{Richard S. Ellis}\affiliation{Department of Physics \& Astronomy, University College London, Gower St., London WC1E 6BT, UK}

\author[0000-0002-5139-4359]{Max Pettini}\affiliation{Institute of Astronomy, Madingley Road, Cambridge CB3 OHA, UK}

\author[0000-0003-2680-005X]{Gabriel Brammer}\affiliation{Niels Bohr Institute, University of Copenhagen, Lyngbyvej 2, DK2100 Copenhagen, Denmark}
\affiliation{Cosmic Dawn Center (DAWN), Copenhagen, Denmark}

\author[0000-0002-3736-476X]{Fergus Cullen}\affiliation{Institute for Astronomy, University of Edinburgh, Royal Observatory, Edinburgh EH9 3HJ, UK}

\author[0000-0003-4264-3381]{Natascha M. F\"{o}rster Schreiber}\affiliation{Max-Planck-Institut f{\"u}r extraterrestrische Physik (MPE), Giessenbachstr. 1, D-85748 Garching, Germany}

\author[0000-0002-0101-336X]{Ali A. Khostovan}\affiliation{Department of Physics and Astronomy, University of Kentucky, 505 Rose Street, Lexington, KY 40506, USA}

\author[0000-0003-4368-3326]{Derek J. McLeod}\affiliation{Institute for Astronomy, University of Edinburgh, Royal Observatory, Edinburgh EH9 3HJ, UK}

\author{Ross J. McLure}\affiliation{Institute for Astronomy, University of Edinburgh, Royal Observatory, Edinburgh EH9 3HJ, UK}

\author[0000-0002-7064-4309]{Desika Narayanan}\affiliation{Department of Astronomy, University of Florida, 211 Bryant Space Sciences Center, Gainesville, FL 32611 USA}
\affiliation{Cosmic Dawn Center at the Niels Bohr Institute, University of Copenhagen and DTU-Space, Technical University of Denmark}

\author[0000-0001-5851-6649]{Pascal A. Oesch}\affiliation{Department of Astronomy, University of Geneva, Chemin Pegasi 51, 1290 Versoix, Switzerland}
\affiliation{Niels Bohr Institute, University of Copenhagen, Lyngbyvej 2, DK2100 Copenhagen, Denmark}
\affiliation{Cosmic Dawn Center (DAWN), Copenhagen, Denmark}

\author[0000-0003-4464-4505]{Anthony J. Pahl}\affiliation{The Observatories of the Carnegie Institution for Science, 813 Santa Barbara Street, Pasadena, CA 91101, USA}

\author[0000-0002-4834-7260]{Charles C. Steidel}\affiliation{Cahill Center for Astronomy and Astrophysics, California Institute of Technology, MS 249-17, Pasadena, CA 91125, USA}

\author[0000-0002-4153-053X]{Danielle A. Berg}\affiliation{Department of Astronomy, The University of Texas at Austin, 2515 Speedway, Stop C1400, Austin, TX 78712, USA}



\begin{abstract}

We present the nebular attenuation curves and dust covering fractions
for 24 $z=1.5-4.4$ star-forming galaxies using multiple Balmer and
Paschen lines from the JWST/AURORA survey.  Nebular reddening derived
from Paschen lines exceeds that from Balmer lines for at least half
the galaxies in the sample when assuming the commonly-adopted Galactic
extinction curve, implying the presence of heavily-reddened star
formation.  The nebular attenuation curves exhibit a broad range of
normalizations ($\rv \simeq 3.2-16.4$).  Motivated by the offsets in
reddening deduced from the Balmer and Paschen lines, and the high
$\rv$ values for the individual nebular attenuation curves, both of
which suggest variations in the dust-stars geometry, we propose a
model with a subunity dust covering fraction ($\fcov$).  Fitting such
a model to the $\hi$ recombination line ratios indicates $\fcov \sim
0.6-1.0$.  The normalizations of the nebular attenuation curves,
$\rv$, are driven primarily by $\fcov$ and the mix of reddened and
unreddened OB associations.  Thus, the diversity of nebular
attenuation curves can be accommodated by assuming dust grain
properties similar to that of Milky Way sightlines but with a subunity
covering fraction of dust.  Integrated measurements of multiple Balmer
and Paschen lines can be used to place novel constraints on the dust
covering fraction towards OB associations.  These, in turn, provide
new avenues for exploring the role of dust and gas covering fraction
in a number of relevant aspects of high-redshift galaxies, including
the impact of stellar feedback on ISM porosity and the escape of
Ly$\alpha$ and Lyman continuum radiation.

\end{abstract}

\keywords{ISM: dust, extinction --- galaxies: evolution --- galaxies: high-redshift --- galaxies: ISM --- galaxies: star formation}

\section{Introduction}
\label{sec:intro}


The accessibility of rest-frame optical and near-infrared
recombination and collisionally-excited emission lines afforded by
recent advances in ground-based near-IR instruments and space-based
missions has motivated detailed modeling of the ISM in distant
star-forming galaxies.  These advances, when combined with ISM studies
of nearby star-forming regions and galaxies, have enabled evolutionary
studies of the state of the ISM throughout most of cosmic history.
Among the most critical of the commonly observed rest-frame optical
and near-infrared emission lines from the ISM are the Balmer and
Paschen series recombination lines of hydrogen.  These lines, in
concert with those from other ionic species, are used to constrain the
chemical abundances, gas densities, temperatures, and the ionization
structure of the ISM.  The hydrogen recombination line ratios
constitute an ideal probe of the dust column density towards $\hii$
regions owing to the cosmic abundance of hydrogen and the
insensitivity of these ratios to gas density and temperature over the ranges typically
inferred for star-forming galaxies.  Thus, these lines are essential
for determining the effect of dust reddening, correcting all nebular
emission lines for wavelength-dependent dust obscuration, and
recovering the intrinsic nebular spectrum of a galaxy.

The $\ha/\hb$ line ratio, or Balmer decrement, is by far the most
commonly employed estimator of nebular dust reddening.  The
constituent lines are the strongest non-resonant recombination
emission lines of hydrogen.  They are simultaneously easily observed
from the ground in certain redshift windows up to $z\sim 2.6$ (e.g.,
\citealt{forster09, kashino13, reddy15}) and now with JWST, up to
$z\sim 6.6$ \citep{shapley23a, sandles24, clarke24}.  They provide
constraints on the dust attenuation towards the most massive stars and
thus complement dust-reddening measures based on non-ionizing UV
continuum emission (i.e., the UV spectral slope,
$\beta$;\citealt{calzetti94, meurer99}) or dust-reprocessed UV (i.e.,
IR) emission, while being less sensitive to star-formation history
\citep{kennicutt94}.  The latest generation of ground-based
multi-object near-IR spectrographs (e.g., Keck/MOSFIRE, the Very Large
Telescope (VLT)/KMOS) and space-based grism capabilities (e.g., the
Hubble Space Telescope (HST)/WFC3 grism) have provided measurements of
Balmer decrements for hundreds of galaxies up to $z\sim 2.6$ (e.g.,
\citealt{forster04, kashino13, dominguez13, price14, reddy15,
  shivaei20b, reddy20, fetherolf21, rezaee21, battisti22, lorenz23}),
allowing for statistical analyses of nebular reddening,
recombination-line-based SFRs, and gas-phase metallicities and
ionization parameters from ratios of lines that are well separated in
wavelength.

Such measurements are now possible well into the epoch of reionization
owing to {\em JWST}'s unprecedented sensitivity, near-IR wavelength
coverage, and spectral resolution \citep{shapley23a, sandles24}, which
allow for the detection of even weaker higher-order Balmer lines
(e.g., $\hg$ up to $z\sim 10.5$, $\hd$ up to $z\sim 11$).  Even
higher-fidelity measurements of nebular dust attenuation (and hence
SFRs, gas-phase metallicities, and ionization parameters) can be
achieved for intermediate-redshift galaxies where the Paschen lines
lie within the wavelength coverage of NIRSpec.
The greater sensitivity and wavelength coverage offered by JWST
naturally benefit studies of a large number of collisionally-excited
and recombination nebular emission lines, allowing us to move beyond
simple strong-line optical diagnostics (e.g., ``BPT'' diagrams;
\citealt{baldwin81}) of the ISM, and perform simultaneous
photoionization modeling of the full suite of rest-frame optical and
near-IR nebular emission lines (e.g., \citealt{steidel16, topping20a,
  reddy22}).  Of course, the simultaneous modeling of many emission
lines that are well separated in wavelength necessitates
wavelength-dependent dust corrections that can be quantified by
calculating the nebular dust attenuation curve.

The far more commonly studied stellar attenuation curve encapsulates
the effects of dust absorption and scattering as determined by the
composition of the dust grains, as well as the scattering of light
into the line of sight, a nonuniform distribution of column densities,
and spatial variations in optical depth within galaxies.  The nebular
attenuation curve captures these effects along the lines of sight
towards the ionized regions,\footnote{The nebular attenuation
  curves are primarily sensitive to the dust foreground to the $\hii$
  regions, as well as any additional contribution of dust within the
  $\hii$ regions (e.g., \citealt{inoue01a}).}  and hence the most
massive stars, in galaxies, while the stellar attenuation curve
generally applies to the stellar continuum contributed by all stars.
Consequently, the stellar and nebular attenuation curves may differ
depending on how the dust is distributed relative to stars of
different masses.  In particular, the main sequence lifetimes of the
most massive O stars (spectral type O6 and earlier;
\citealt{leitherer90}) are shorter than the typical molecular cloud
crossing timescale of $\la 20$\,Myr \citep{calzetti94}.  Thus, while
these stars are on the main sequence and dominating the ionizing flux,
they are still embedded within their birth clouds and are found in
regions of higher dust column density (c.f., \citealt{conroy12}).
Consequently, differences between the stellar and nebular dust
attenuation curves may be expected if there are variations in dust
grain size, absorption and scattering properties, spatial
distribution, or column density between sightlines to massive O stars
and sightlines to lower-mass stars contributing significantly to the
non-ionizing UV continuum flux.

For starburst galaxies at high redshift, the \citet{calzetti00} and
Small Magellanic Cloud (SMC; \citealt{gordon03}) curves are commonly
assumed to describe the reddening of the stellar continuum, while the
reddening towards the nebular regions is generally assumed to be
described by the Galactic extinction curve \citep{cardelli89}.
Constraining
the stellar attenuation curve is generally more complex relative to
the nebular attenuation curve.  The former is highly model dependent,
relying on assumptions about the intrinsic SED, which can vary
considerably depending on the specifics of stellar population
synthesis models and the star-formation history.  In contrast, the
nebular attenuation curve can be more straightforwardly constrained
through photoionization modeling to deduce intrinsic $\hi$
recombination line ratios.

Using optical, mid-IR, and radio $\hi$ recombination lines detected
for the local starburst galaxy, M82, \citet{forster01} find a nebular
reddening curve inconsistent with a foreground screen.  They find that
a mixed dust-stars model provides the best fit to the recombination
lines.  Using multiple Balmer emission lines ($\ha$, $\hb$, $\hg$,
$\hd$, and $\he$) detected in ground-based composite spectra of a
representative sample of $z\sim 2$ star-forming galaxies,
\citet{reddy20} placed the first direct constraint on the nebular
attenuation curve at high redshift (see \citealt{rezaee21} for a
similar analysis of nearby star-forming galaxies).  They demonstrated
that, on average, the shape of the curve at rest-frame optical
wavelengths is similar to that of the Galactic extinction curve
\citep{cardelli89}, the most commonly assumed curve describing nebular
reddening \citep{calzetti94}.  \citet{prescott22} used a combination
of ground-based Balmer decrements and HST-based grism measurements of
Pa$\beta$ for 11 galaxies at $z<0.3$ to find a diversity of
attenuation curve slopes and normalizations.  These initial efforts to
constrain the nebular attenuation curve underscored the need for deep
observations over a wide baseline in wavelength, with
well-characterized corrections for aperture (or slit) losses and/or
relative flux calibration between different observational
configurations to ensure accurate line ratios.

Initial surveys with JWST---e.g., the Cosmic Evolution Early Release
Science survey (CEERS, \citealt{finkelstein23}), and the JWST Advanced
Deep Extragalactic Survey (JADES, \citealt{eisenstein25})---have
demonstrated the transformative improvement in sensitivity, wavelength
coverage, and spectral resolution necessary for {\em simultaneous}
detections of the full range of Balmer and Paschen-series
recombination emission lines for intermediate-redshift galaxies at
$z\sim 1 - 4$.  These surveys have enabled the first statistical
analysis of nebular reddening and SFRs based on both Balmer and
Paschen lines for the same galaxies at these redshifts
\citep{reddy23a}.  Now, with considerably higher-S/N JWST/NIRSpec
spectra collected for statistical samples of galaxies at intermediate
redshifts, we can provide the most stringent constraints to date on
the shape and normalization of the nebular attenuation curve for
individual galaxies.  Aside from allowing accurate nebular dust
corrections on an individual object-by-object basis, these datasets
enable us to examine the scatter in the nebular attenuation curve and
its effect on other commonly-derived quantities such as ionization
parameters, metallicities, and star-formation rates.

Here we take advantage of the deep near-IR spectroscopy provided by
the Cycle 1 JWST/NIRSpec program ``The Assembly of Ultradeep
Observations Revealing Astrophysics'' (AURORA, PID:1914).  This survey
was designed to detect faint auroral emission lines from ionized O, S,
and N in $z>1.4$ galaxies, with the aim of measuring electron
temperatures and obtaining direct estimates of O abundance,
independent of ``strong-line'' estimates of gas-phase metallicity
which suffer from a number of systematic uncertainties.  As a result
of the required depth to detect the weak auroral lines, AURORA spectra
represent a marked improvement in $S/N$ for typical star-forming
galaxies at intermediate redshifts compared to earlier surveys (e.g.,
CEERS).  In particular, these spectra include the robust detection of
many of the weaker higher-order Balmer and Paschen-series emission
lines over the continuous wavelength range of $1-5$\,$\mu$m, along
with significant S/N in the stellar continuum that aids in the
relative flux calibration between gratings, thus obviating much of the
systematic uncertainty in flux calibration between measurements
obtained with different telescopes and instruments and in different
weather conditions (e.g., \citealt{prescott22}).  Using the AURORA
spectroscopy, \citet{sanders25} adopt the methodology laid out in
\citet{reddy20} to constrain the shape of the nebular dust attenuation
curve for a young $z=4.411$ galaxy, GOODSN-17940, finding a curve that
deviates significantly in shape from the Galactic extinction curve.
Fully quantifying the diversity in the shape and normalization of the
nebular dust attenuation curve requires a statistical analysis of
objects spanning a range of galaxy properties (e.g., SFR, stellar
mass, gas-phase abundance) that are known to correlate with dust
attenuation.  To that end, in this paper we use the deep NIRSpec
spectroscopy from the AURORA survey to provide the first robust
constraints on the nebular reddening and the nebular attenuation curve
for a statistical sample of individual galaxies based on multiple
Balmer and Paschen emission lines.  We further demonstrate how the
detection of multiple HI recombination lines can provide novel
constraints on the dust covering fraction toward OB associations.

The paper is organized as follows.  The AURORA survey, data reduction,
slitloss corrections, primary measurements, and sample selection are
outlined in Section~\ref{sec:data}.  The comparison between the
Balmer-inferred and Paschen-inferred nebular reddening is presented in
Section~\ref{sec:nebred}.  Section~\ref{sec:effcurve} builds on
previous work (e.g., \citealt{reddy20, rezaee21}) by focusing on the
derivation of nebular dust attenuation curves for individual galaxies.
Finally, Section~\ref{sec:discussion} explores a more sophisticated
model that allows for a sub-unity covering fraction of dust in order
to simultaneously reproduce the observed $\hi$ Balmer and Paschen
recombination line ratios.  The broader implications of the nebular
dust attenuation curves in terms of dust-corrected line luminosities
and line ratios, SFRs, and differential reddening of the nebular lines
and stellar continuum, are presented in a separate paper, hereafter
Paper II (Reddy et~al., submitted).  A \citet{chabrier03} initial mass
function (IMF) is considered throughout the paper.  Wavelengths are
reported in the vacuum frame.  We adopt a cosmology with
$H_{0}=70$\,km\,s$^{-1}$\,Mpc$^{-1}$, $\Omega_{\Lambda}=0.7$, and
$\Omega_{\rm m}=0.3$.

\section{Data and Measurements}
\label{sec:data}

\subsection{Observations and Spectral Data Reduction}

The JWST/NIRSpec spectroscopy used in this analysis was drawn from
the AURORA survey (PID: 1914, co-PIs: Shapley and Sanders).  The
central strategy of the survey was to target the auroral lines of
multiple elements (O, N, and S) for individual galaxies at $z>1.4$,
with the aim of constraining their electron temperatures and obtaining
direct O abundances.  The survey targeted 46 and 51 galaxies,
respectively, in the COSMOS and GOODS-N fields.  Details of the target
selection are provided in \citet{shapley25}.  The G140M/F100LP,
G235M/F170LP, and G395M/F290LP grating/filter combinations were used,
providing $R\sim 1000$ spectroscopy over the full wavelength range
spanning 1 to 5\,$\mu$m.  A three-dither nodding pattern was used and
the total integration times in the three grating/filter combinations
for each pointing were 12.3, 8.0, and 4.2\,hrs, respectively, yielding
a uniform $3\sigma$ line flux detection limit of $5\times
10^{-19}$\,erg\,s$^{-1}$\,cm$^{-2}$ from 1 to 5\,$\mu$m.

The standard STScI NIRSpec data reduction pipeline and custom software
developed for the AURORA survey were used to process the raw data and
obtain flat-fielded and wavelength-calibrated two-dimensional (2D)
spectrograms.  Optimal extraction was used to create one-dimensional
(1D) science and error spectra from the 2D spectrograms, using a
spatial profile determined by the brightest emission line in each
grating, or the integrated continuum profile if no emission lines
were detected.  Only for $\sim 1\%$ of the targets was a blind extraction
performed.  Further details on the survey parameters, including target
selection and data reduction, can be found in \citet{shapley25}.

\subsection{Slit Loss Corrections and Flux Calibration}
\label{sec:slitloss}

Owing to the small spatial scale of the NIRSpec MSA microshutter
($0\farcs 20 \times 0\farcs 46$) relative to the typical size of the
targeted galaxies, a substantial fraction of the light from these
galaxies falls outside the spectroscopic aperture.  This issue is
particularly acute for galaxies significantly offset from the center
of the microshutter.  To correct for the light lost outside the
aperture, one must account for the aperture size, the intrinsic size
of the galaxy, any offset of the galaxy within the aperture, and the
wavelength-dependent point-spread function (PSF).

The methodology used to correct for this ``slit loss'' follows that of
\citet{reddy23a}.  Specifically, a $12\arcsec \times 12\arcsec$
subimage centered on the target was extracted from the NIRCam F115W
imaging.  Using the segmentation map, pixels corresponding to
unrelated objects were masked.  The subimage was then rotated to
account for the position angle of the NIRSpec observations, and
convolved with Gaussian kernels to simulate the expected light
profiles at longer wavelengths.  These kernels were calculated by
subtracting in quadrature the FWHM of the JWST PSF at $1.15$\,$\mu$m
from the FHWMs of the JWST PSFs at $\lambda = 1.2-5.3$\,$\mu$m, in
increments of $\Delta\lambda =0.1$\,$\mu$m.

For four objects in COSMOS and two in GOODS-N that lacked NIRCAM F115W
imaging, S\'ersic fits from the HST/WFC3 F160W images, as presented in
\citep{vanderwel14}, were used to model the galaxies.  These models
were then convolved with the JWST PSFs at $\lambda = 0.6-5.3$\,$\mu$m,
in increments of $\Delta\lambda = 0.1$\,$\mu$m, generated using the
JWST WebbPSF
software.\footnote{https://www.stsci.edu/jwst/science-planning/proposal-planning-toolbox/psf-simulation-tool}
Additionally, for five targets in GOODS-N, which did not have reliable
S\'ersic fits or F115W imaging, a point-source model was assumed.

The above procedure generates convolved images of the galaxy at each
wavelength.  These images were shifted according to the target's
offset from the microshutter center and then masked.  The masking
accounts for the $0\farcs 20$ microshutter width, the $0\farcs 07$ gap
between adjacent microshutters along the cross-dispersion axis, and
the window used extract the 1D spectra from the 2D spectrograms.  The
1D spectra were corrected for slitloss by dividing them by the
fraction of light transmitted within the spectroscopic and extraction
apertures as a function of wavelength.

\begin{figure}
  \epsscale{1.2}
  \includegraphics[width=1.0\linewidth]{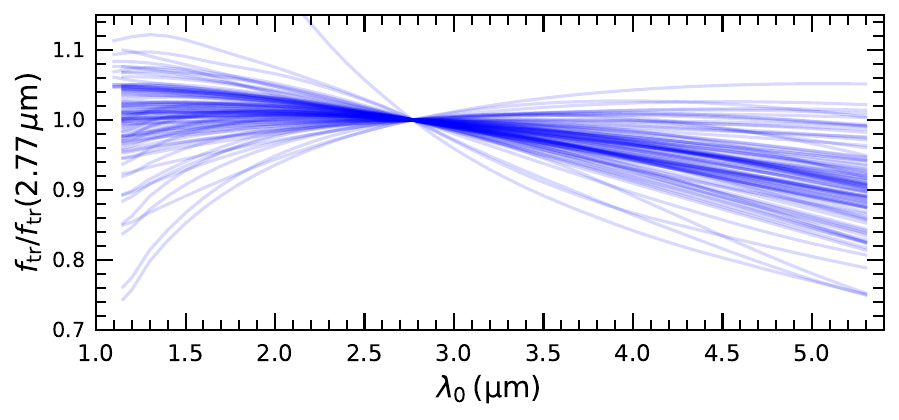}
    \caption{Fraction of transmitted flux as a function of wavelength,
      normalized by the fraction of transmitted flux at
      $2.77$\,$\mu$m, for the 96 galaxies in the AURORA parent
      sample.}
    \label{fig:slitloss}
\end{figure}

The impact of the slitloss corrections are illustrated in
Figure~\ref{fig:slitloss}, which shows that the fraction of
transmitted light as a function of wavelength typically decreases by
$\simeq 10\%$ from $\lambda_0 = 1$ to $5$\,$\mu$m.  However, for some
galaxies that contain subcomponents that are substantially off-center,
the widening of the PSF can increase the fraction of transmitted flux
at longer wavelengths.

The final flux calibration of the spectra involved two steps.  First,
the G140M and G395M spectra were scaled using the G235M grating
spectra as a reference, ensuring that the line fluxes and continuum
flux densities in overlapping wavelength regions matched.  This step
is facilitated by the deep AURORA spectra, which provide high $S/N$
detections of the continua for all galaxies relevant to our analysis.
Second, an absolute flux calibration was performed by scaling the
spectra to match the available multi-wavelength photometry.  Further
details on the relative and absolute flux calibrations can be found in
\citet{sanders25}.

\subsection{Line Measurements and Stellar Photospheric Absorption Corrections}
\label{sec:linemeas}

Emission-line flux measurements were performed by fitting a single
Gaussian function to well-isolated lines, or multiple Gaussians for
lines closely separated in wavelength.  In fitting the lines, the
continuum was initially defined by the best-fit SED model
(Section~\ref{sec:sedfitting}) to the multi-wavelength photometry,
prior to the application of any corrections for emission lines or
nebular continuum.  The resulting emission-line measurements were then
used to correct the multi-wavelength photometry.  The continuum from
the best-fit SED model to this corrected photometry was subsequently
used to refit all the lines and obtain the final flux measurements.
Because the continuum is defined by the best-fit SED model, the final
flux measurements for the $\hi$ recombination lines account for
underlying stellar photospheric absorption.

The impact of stellar absorption on the measured line fluxes was
  evaluated by remeasuring all lines assuming a linear continuum with
  no absorption component.  Due to slight variations in the continuum
  fitting, these resulting line fluxes can occasionally exceed those
  measured with stellar absorption included.  In some cases,
  particularly for the weaker higher-order Balmer lines, a linear
  continuum fit yields no measurable emission.  For the Balmer lines,
  the median corrected-to-uncorrected line ratios vary from $\simeq
  1.78$ for H12 to $1.01$ for $\ha$.  Similarly, for the Paschen lines,
  the median corrected-to-uncorrected line ratios vary from $1.36$ for
  Pa12 to $1.01$ for Pa4.

Slight systematic variations in the predicted stellar absorption arise
depending on the assumed stellar population synthesis (SPS) model and
star-formation history.  For example, continuum fits based on the
Binary Population and Spectral Synthesis (BPASS) version 2.2.1
constant star-formation (CSF) models \citep{eldridge17, stanway18}
yield absorption-corrected line fluxes that are $\simeq 20\%$ to
$<1\%$ higher than those obtained with the flexible ``delayed-$\tau$''
star-formation histories adopted here (see
Section~\ref{sec:sedfitting}).  The largest discrepancies occur for
the three highest-order Balmer lines analyzed here (H10, H11, and
H12), while the corrections are minimal ($\la 5\%$) for the
lower-order Balmer lines and $\la 1\%$ for $\ha$.  Likewise, the
Paschen line fluxes are $\simeq 15\%$ to $\la 1\%$ higher assuming the
CSF models, with larger corrections for the higher-order (and weaker)
Paschen lines.  The effect of these stellar absorption corrections on
the various parameters derived in this work are discussed further in
Sections~\ref{sec:indandave}, \ref{sec:rvnorm}, and
\ref{sec:covfracparms}.  We note that trends analyzed here and the
main conclusions are unaffected by the choice of SPS model used to
correct for stellar absorption.


\subsection{SED Fitting}
\label{sec:sedfitting}

The COSMOS and GOODS-N fields contain publicly available photometry
from JWST/NIRCam and HST/ACS and WFC3, which were obtained from the
Dawn JWST Archive (DJA; \citealt{valentino23, heintz25}).  In COSMOS,
the bands include the following: F435W, F606W, F814W, F850LP, F105W,
F125W, F140W, and F160W from HST; and F090W, F115W, F150W, F200W,
F277W, F356W, F410M, and F444W from JWST/NIRCam.  In GOODS-N, the
bands include the following: F435W, F606W, F775W, F814W, F850LP,
F105W, F125W, F140W, and F160W from HST; and F090W, F115W, F150W,
F182M, F200W, F210M, F277W, F335M, F356W, F410M, and F444W from
JWST/NIRCam.  For four sources in COSMOS that do not contain
JWST/NIRCam imaging, photometry from the 3D-HST catalog was used
\citep{skelton14}.  Details on how the photometry was performed are
given in \citet{valentino23}.

The photometry was used to model the stellar populations of the
galaxies, yielding estimates of SFRs, ages, continuum reddening
($\ebmvcont$), and stellar masses.  The \citet{conroy09} flexible
  ``delayed-$\tau$'' models were assumed, with either the SMC or
  Calzetti dust attenuation curves depending on the stellar mass and
  redshift of the galaxy (see \citealt{shapley25} for further
  details).
For the purposes of this analysis, the best-fit SEDs, are only used
for measuring emission-line fluxes (Section~\ref{sec:linemeas}) and
estimating continuum reddening (Section~\ref{sec:diffuse}).  Paper II
also considers the BPASS CSF models, and provides a detailed analysis
of the best-fit SED parameters.


\begin{figure*}
  \epsscale{1.2}
  \includegraphics[width=1.0\linewidth]{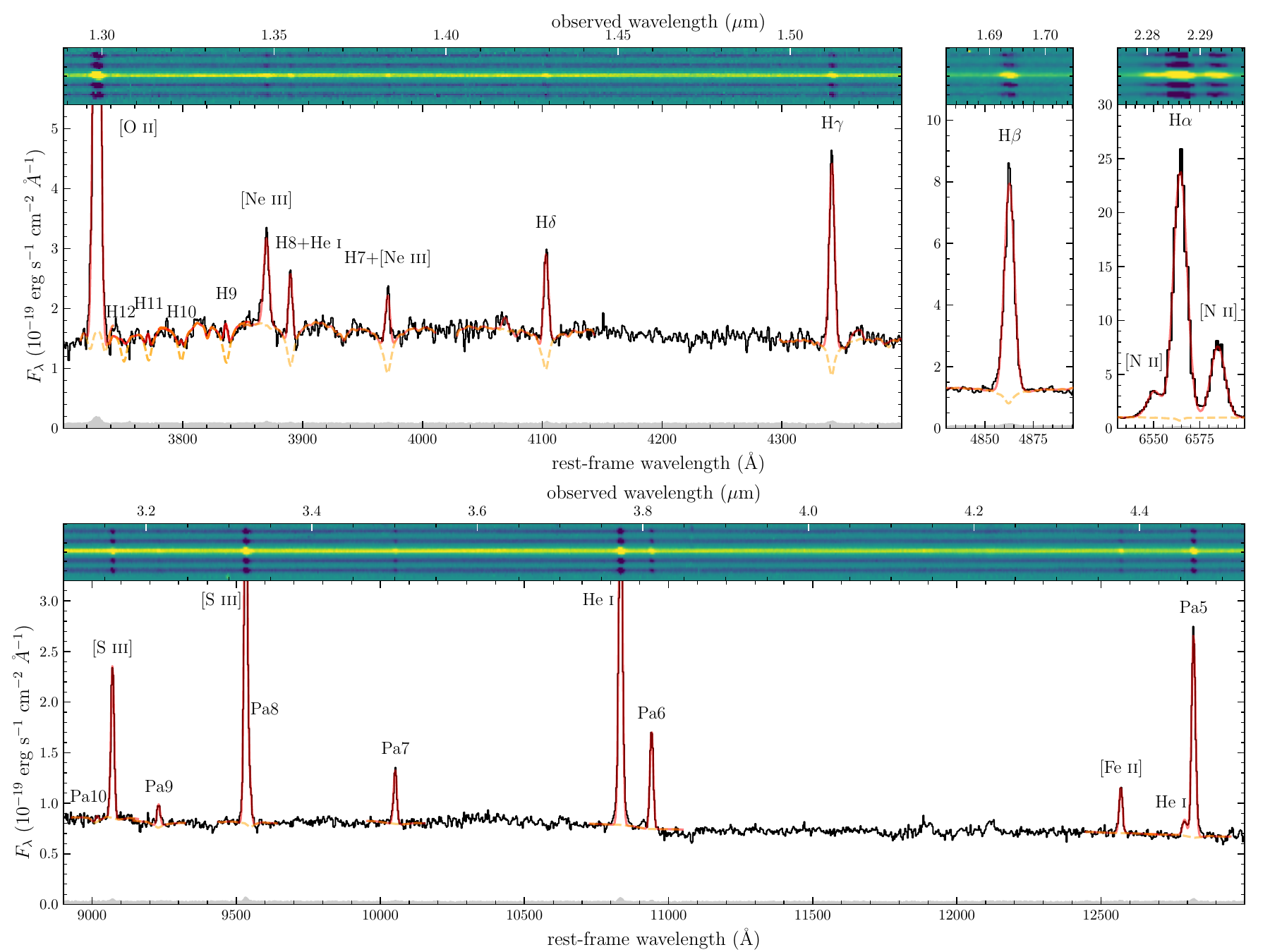}
    \caption{Two- and one-dimensional JWST/NIRSpec spectra of one of
      the objects in the main sample, GOODSN-30564 ($z=2.4828$),
      showing the detection of multiple $\hi$ Balmer and Paschen
      emission lines.  The 1D spectrum is shown in black, and the
      error spectrum is indicated in grey.  The underlying stellar
      continuum and fits to the emission lines are shown in orange and
      red, respectively.}
    \label{fig:examplespectrum}
\end{figure*}

\subsection{Final Sample Selection}
\label{sec:sampleselection}

\begin{figure}
  \epsscale{1.0}
  \includegraphics[width=1.0\linewidth]{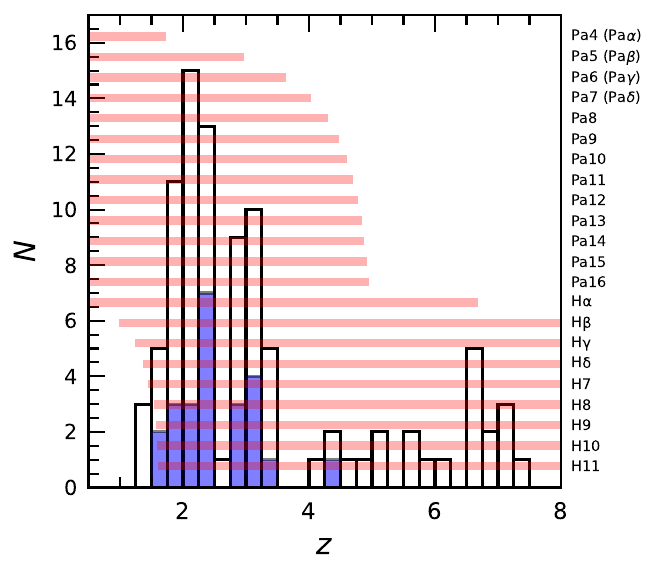}
    \caption{Redshift distribution of the AURORA sample (open
      histogram), with the final sample of 24 galaxies indicated by
      the blue shaded histogram.  The line coverage of the sample is
      demonstrated by the horizontal lines which indicate the redshift
      range over which the Balmer and Paschen lines lie within the
      wavelength coverage of JWST/NIRSpec.}
    \label{fig:redshifts}
\end{figure}

Of the 94 galaxies with spectroscopic redshifts in the parent AURORA
sample, we culled a subsample suitable for the present analysis.
First, 8 AGN and quiescent galaxies were excluded based on prior known
information for these sources (see \citealt{shapley25}), and
inspection of the SEDs and/or AURORA spectra for signatures of broad
lines or lack of emission lines.  Second, 9 objects where the flux
calibration was suspect due to inconsistent line ratios between
adjacent gratings or inconsistent/unphysical line ratios within the
same grating were removed.  Third, 2 galaxies exhibiting observed line
ratios inconsistent with Case B recombination were removed.  Fourth,
to ensure robust constraints on the nebular dust attenuation curve, 4
galaxies which have a negligible reddening based on the Balmer lines
were removed.  Fifth, we removed 34 galaxies at higher redshifts which
did not have coverage of any Paschen lines.  Finally, we required
$>5\sigma$ detections of at least 5 lines, at least 2 of which must be
Paschen lines, resulting in the exclusion of 13 additional galaxies.
Applying the above criteria yielded a sample of 24 galaxies, each
having anywhere from 5 to 16 $5\sigma$ detections of the $\hi$ Balmer
and Paschen recombination lines, with a median of 11 lines.
Figure~\ref{fig:examplespectrum} shows the 2D and 1D spectra of one of
the objects in the sample, GOODSN-30564, which has 9 detected Balmer
lines and 5 detected Paschen lines.  The redshift distributions of the
parent AURORA sample and the final sample of 24 galaxies used for this
analysis are shown in Figure~\ref{fig:redshifts}.  In addition to the
24 galaxies in the primary sample, we also consider in
Section~\ref{sec:highzcurve}
additional galaxies, many at $z\ga 4$, that did not
have more than two detected Paschen lines.

%

\begin{figure}
  \epsscale{1.2}
  \includegraphics[width=1.0\linewidth]{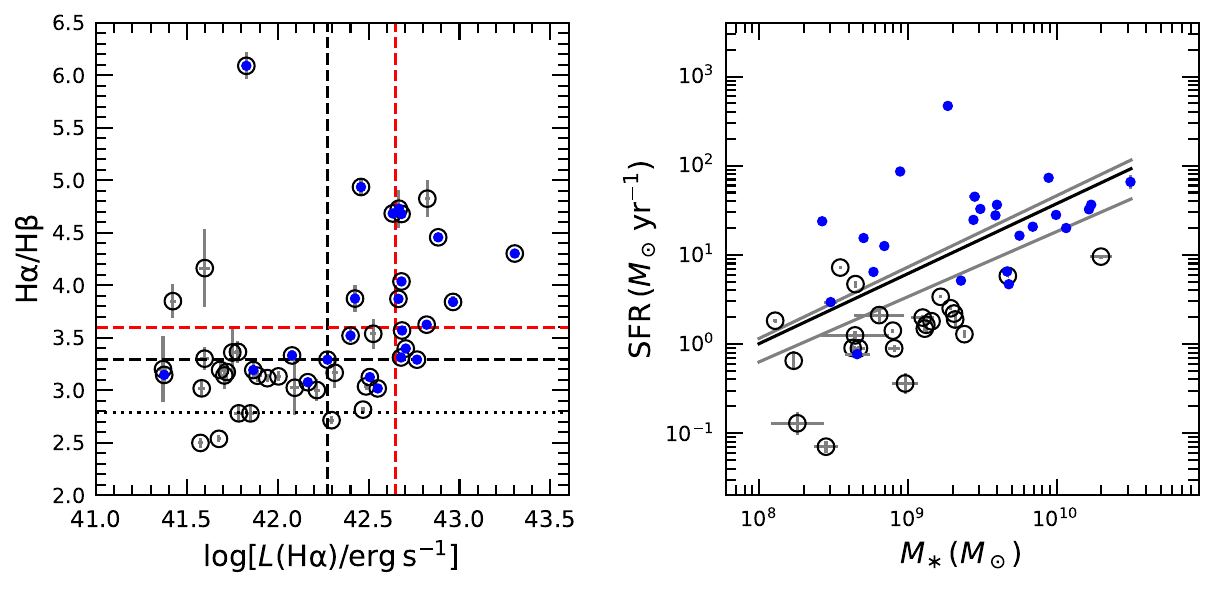}
    \caption{{\em Left:} Distributions of the observed $\ha$
      luminosities and Balmer decrements for the parent AURORA sample
      with $>5$\,$\sigma$ detections of $\ha$ and $\hb$ (open
      symbols).  The subsample of 24 galaxies included in this
      analysis are indicated by the blue symbols.  The dashed black
      vertical and horizontal lines indicate the median $\ha$
      luminosity and Balmer decrement, respectively, for the parent
      sample, and the dashed red vertical and horizontal lines
      indicate the same for the subsample of 24 galaxies.  The dotted
      black line indicates the intrinsic $\ha/\hb$ ratio of $2.788$.
      {\em Right:} Relation between SFR and $M^\ast$ for the 24
      galaxies analyzed here (blue symbols), along with additional
      AURORA galaxies at similar redshifts $z=1.5-4.4$ which did not
      satisfy the selection criteria discussed in
      Section~\ref{sec:sampleselection} (open symbols).  Also shown
      are the parameterized SFR-$M^\ast$ relations at $z=1.5$, $z=3$,
      and $z=4.4$ from \citet{speagle14} (lines).}
    \label{fig:hahb}
\end{figure}

Figure~\ref{fig:hahb} shows the distributions of the observed $\ha$
luminosities and Balmer decrements for the parent AURORA sample with
$>5\sigma$ detections of $\ha$ and $\hb$, along with the same
distributions for the subsample of 24 galaxies that satisfy the
aforementioned selection criteria.  Not surprisingly, the median
Balmer decrement is higher for the subsample of 24 galaxies ($\ha/\hb
\simeq 3.6$ vs $3.3$ for the parent sample)---translating into a
  difference in reddening of $\Delta\ebmvneb = 0.088$ assuming the
  \citet{cardelli89} curve---since we only included galaxies where
the Balmer lines indicate non-negligible reddening.  The additional
requirement of the detection of at least 2 Paschen lines with $S/N >
5$ results in a subsample with a median observed $\ha$ luminosity that
is $\simeq 0.4$\,dex larger than that of the parent sample.  Also
  shown in Figure~\ref{fig:hahb} is the distribution of the SFRs and
  stellar masses of the subsample of 24 galaxies relative to the
  remaining galaxies in the AURORA sample at redshifts
  $z=1.5-4.4$ that fell out of the sample.  The 24 galaxies analyzed
  here generally follow the SFR-$M^\ast$ relations established for
  galaxies at similar redshifts \citep{speagle14}, though they tend to
  be more massive and have higher SFRs relative to galaxies that fell
  out of the sample.  Again, these offsets are not surpising given
  that the 24 galaxies analyzed here were selected to have
  non-neglible Balmer reddening, corresponding to dustier galaxies
  with higher SFRs and stellar masses.


\section{Nebular Reddening from the Balmer and Paschen Lines}
\label{sec:nebred}

The longer-wavelength Paschen lines are less affected by dust
attenuation than the Balmer lines and can thus provide an independent
probe of the intrinsic $\hi$ recombination line luminosities.  As we
show below, the combination of the Balmer and Paschen lines provides
the most stringent constraints on not only the nebular reddening, but
also the shape and normalization of the nebular attenuation curve.
For the moment, we focus on the nebular reddening derived from the
Balmer and Paschen line ratios adopting the commonly-assumed Galactic
extinction curve of \citet{cardelli89}.  The observed flux of a line
at wavelength $\lambda$, $f(\lambda)$, can be expressed as a function
of the intrinsic flux of that line, $f_{\rm 0}(\lambda)$, the nebular
reddening, $\ebmvneb$, and the nebular attenuation curve, $k_{\rm
  neb}$:
\begin{equation}
f(\lambda) = f_0(\lambda) \times 10^{-0.4\ebmvneb k_{\rm neb}(\lambda)}.
\label{eq:eq1}
\end{equation}
The ratio of the observed fluxes of two lines at wavelengths $\lambda_1$ and $\lambda_2$ can then be written as
\begin{equation}
\frac{f(\lambda_1)}{f(\lambda_2)} = \frac{f_0(\lambda_1)}{f_0(\lambda_2)}\times 10^{-0.4\ebmvneb[k_{\rm neb}(\lambda_1) - k_{\rm neb}(\lambda_2)]}.
\label{eq:eq2}
\end{equation}
Defining 
\begin{equation}
R\equiv \log_{10}\left[\frac{f(\lambda_1)}{f(\lambda_2)}\right] - \log_{10}\left[\frac{f_0(\lambda_1)}{f_0(\lambda_2)}\right],
\label{eq:eq3}
\end{equation}
we can write
\begin{equation}
R = -0.4\ebmvneb [k_{\rm neb}(\lambda_1) - k_{\rm neb}(\lambda_2)].
\label{eq:eq4}
\end{equation}
The observed and intrinsic line ratios and an assumed nebular
attenuation curve can then be used to calculate the nebular reddening
using Equations~\ref{eq:eq3} and \ref{eq:eq4}.  

Intrinsic $\hi$ recombination line ratios were obtained from PyNeb
\citep{luridiana15} with an electron density $n_e = 100$\,cm$^{-3}$
and temperature $T=15,000$\,K.  This $n_e$ is within a factor of
$\simeq 3$ of the typical densities found in galaxies at $z\sim 2-3$,
as indicated by both ground-based spectroscopy of samples spanning
similar redshifts (e.g., \citealt{steidel14, steidel16, sanders16a,
  strom17, topping20a, reddy23b}) and, more recently, with JWST
spectroscopy of galaxies at the same redshifts (e.g.,
\citealt{reddy23c, isobe23, topping25}).  The impact of varying $n_e$
within the range inferred for star-forming galaxies ($n_e <
10^4$\,cm$^{-3}$) on the intrinsic line ratios is negligible.
However, a factor of $\approx 2$ variation in temperature at a fixed
$n_e$ can cause non-negligible shifts in line ratios, particularly for
the higher-order Balmer lines and the Paschen lines (e.g.,
\citealt{ferguson97}).  Appendix~\ref{sec:intrinsic} summarizes the
variations in the line ratios obtained at different temperatures.  The
appendix further presents different treatments of
angular-momentum-state mixing (or $l$-mixing) in various principle
quantum ($n$) states and the impact on the intrinsic line ratios,
which ultimately influences the derived shape of the nebular dust
attenuation curve.  For ease of reference, the default PyNeb values
mentioned above yield the intrinsic line ratios listed in
Table~\ref{tab:intrinsic}, which are adopted for the subsequent
analysis.




\begin{deluxetable}{lrc}
\tabletypesize{\footnotesize}
\tablewidth{0pc}
\tablecaption{Intrinsic $\hi$ Recombination Line Emissivities Relative to $\ha$}
\tablehead{
\colhead{Line} &
\colhead{$\lambda$ (\AA)\tablenotemark{a}} & 
\colhead{$I$\tablenotemark{b}}}
\startdata
H12 &  3751.22 & 0.01106 \\
H11 &  3771.70 & 0.01440 \\
H10 &  3798.98 & 0.01925 \\
H9  &  3836.48 & 0.02657 \\
H8\tablenotemark{c}  &  3890.17 & 0.03819 \\
H7\tablenotemark{d}  &  3971.20 & 0.05783 \\
$\hd$ &  4102.89 & 0.0941 \\
$\hg$ & 4341.68 & 0.16960 \\
$\hb$ & 4862.68 & 0.35863 \\
$\ha$ & 6564.61 & 1.00000 \\
Pa16 &  8504.83 & 0.00154 \\
Pa15 &  8547.73 & 0.00186 \\
Pa14 &  8600.75 & 0.00229 \\
Pa13 &  8667.40 & 0.00287 \\
Pa12 &  8752.86 & 0.00365 \\
Pa11 &  8865.32 & 0.00475 \\
Pa10 &  9017.77 & 0.00634 \\
Pa9 &  9232.23 & 0.00875 \\
Pa8 &  9548.82 & 0.01255 \\
Pa7 & 10052.56 & 0.01897 \\
Pa6 & 10941.17 & 0.03070 \\
Pa5 & 12821.58 & 0.05459 \\
Pa4 & 18756.42 & 0.10930 \\
\enddata
\tablenotetext{a}{Rest-frame vacuum wavelength.}
\tablenotetext{b}{Intensity of line relative to $\ha$ for Case B recombination, $T_{\rm e}=15,000$\,K,
and $n_{\rm e}=100$\,cm$^{-3}$, taken from PyNeb.}
\tablenotetext{c}{H8 is blended with $\hei$\,$\lambda 3890$ and was not used in the analysis.} 
\tablenotetext{d}{H7 ($\he$) is blended with $\neiii\lambda 3969$.  The contribution of the latter to the total line flux 
was calculated by multiplying the measured $\neiii\lambda 3870$ flux by 0.31 (Section~\ref{sec:linemeas}).}
\label{tab:intrinsic}
\end{deluxetable}

\begin{figure}
  \epsscale{1.2}
  \includegraphics[width=1.0\linewidth]{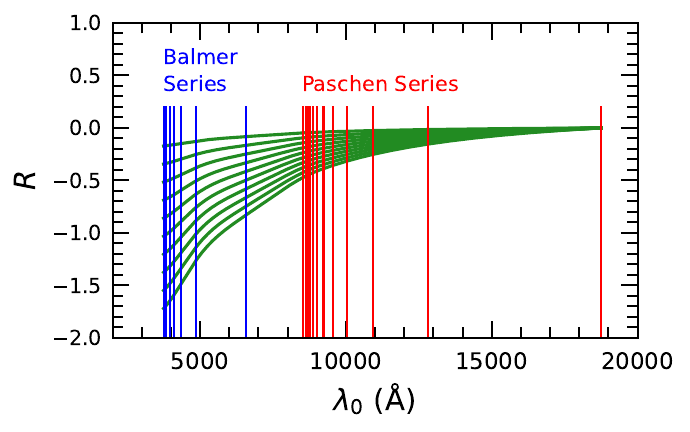}
    \caption{$R$ versus rest-frame wavelength for $\ebmvneb = 0.1-1.0$
      in steps of $\delta\ebmvneb = 0.1$ (proceeding from top to
      bottom) assuming the \citet{cardelli89} extinction curve (green
      lines).  All curves have been normalized relative to the
      attenuation of Pa4.  The wavelengths of the Balmer and Paschen
      series lines are indicated by the blue and red vertical lines,
      respectively.}
    \label{fig:ebmvdemo}
\end{figure}

To get a sense for how robustly $\ebmvneb$ can be constrained with the
Balmer and Paschen lines, we show in Figure~\ref{fig:ebmvdemo} $R$
versus rest-frame wavelength ($\lambda_0$) for $\ebmvneb = 0.1 - 1.0$
in increments of $\delta\ebmvneb = 0.1$ assuming the
\citet{cardelli89} curve, where the reference wavelength (i.e.,
$\lambda_2$ in Equation~\ref{eq:eq3}) is taken to be that of Pa4
(Pa$\alpha$).  Due to the shallower wavelength dependence of the
\citet{cardelli89} curve, and attenuation curves in general, with
increasing wavelength, the longer-wavelength Paschen lines provide
looser constraints on the nebular reddening compared to the
shorter-wavelength Balmer lines.  For a given uncertainty in the
measured line ratios, the Balmer lines (or a combination of Balmer and
Paschen lines) allow for a better discrimination of $\ebmvneb$ as the
shorter wavelength lines are more sensitive to dust reddening.
Furthermore, because the Balmer lines are intrinsically stronger than
the Paschen lines---and, even when reddened, the observed Balmer line
emission is typically stronger than the observed Paschen line
emission---the higher $S/N$ Balmer lines provide additional leverage
on $\ebmvneb$.  Thus, the strength of the Balmer lines and their
sensitivity to reddening result in tighter constraints on $\ebmvneb$
than what can be achieved with the Paschen lines alone.  The depth of
the AURORA spectroscopic data is such that high signal-to-noise is
achieved even in the higher-order Paschen lines ($S/N>5$ for the
galaxies considered here), allowing for a probe of nebular reddening
independent of the Balmer lines, albeit with larger uncertainties.

\begin{figure*}
  \epsscale{1.2}
  \includegraphics[width=1.0\linewidth]{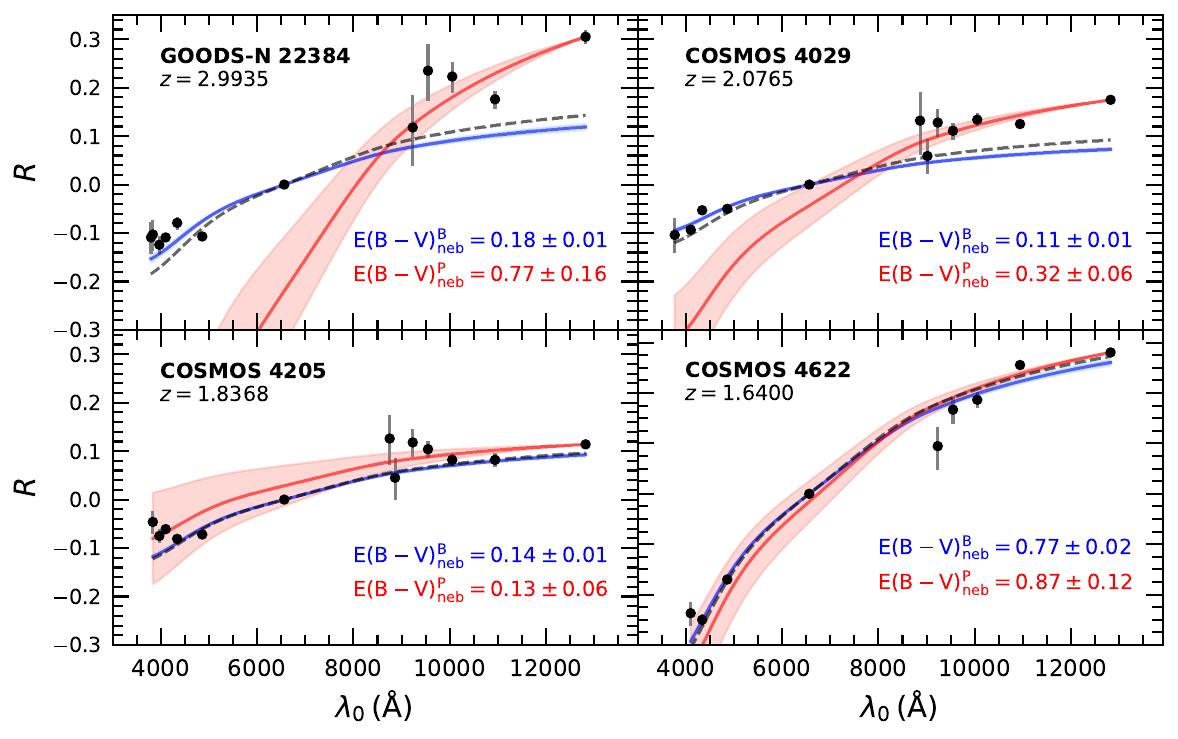}
    \caption{$R$ versus rest-frame wavelength for a representative
      subsample of 6 galaxies.  The measured values and uncertainties
      in $R$ are indicated by the black points and errorbars.  The
      predicted relationship between $R$ and $\lambda_0$ for the
      $\ebmvneb$ that best fits the Balmer lines only and the Paschen
      lines only, are shown by the blue and red curves, respectively,
      while the shaded regions indicate the $68\%$ confidence
      intervals in the fits.  The dashed lines indicate the same when
      fitting all available lines.  $R$ is computed with $\ha$ as the
      reference line, and the predictions shown here assume the
      \citet{cardelli89} extinction curve.  The best-fit
      $\ebmvneb^{\rm B}$ and $\ebmvneb^{\rm P}$ are indicated in each
      panel.  
The sample includes a
      mix of galaxies, some with comparable Balmer and Paschen
      inferred reddening, and some where the Paschen inferred reddening
      exceeds that from the Balmer lines.}
    \label{fig:examples}
\end{figure*}

Figure~\ref{fig:examples} shows $R$ versus $\lambda_0$ for a
representative subsample of galaxies, where $\ha$ is chosen as the
reference line.  In each panel, the blue line and blue shaded region
indicate the relationship and $1\sigma$ uncertainty between $R$ and
$\lambda_0$ for the $\ebmvneb$ that best fits the Balmer lines,
$\ebmvneb^{\rm B}$, adopting the \citet{cardelli89} extinction curve.
The best-fit $\ebmvneb$ was determined using $\chi^2$ fitting.
Similarly, the red line and red shaded region indicate the
relationship and $1\sigma$ uncertainty between $R$ and $\lambda_0$ for
the $\ebmvneb$ that best fits the Paschen lines, $\ebmvneb^{\rm P}$.
As discussed above, the uncertainty in $\ebmvneb$ obtained from the
Paschen lines alone is generally larger than that obtained from the
Balmer lines.  The dashed lines in each panel indicate the fit to all
available lines, which, due to the high S/N (and relatively high
weighting) of the Balmer lines, yields an $\ebmvneb$ that is close to
$\ebmvneb^{\rm B}$.  The top two panels of Figure~\ref{fig:examples}
show two galaxies where the $\ebmvneb$ obtained from the Balmer lines
alone is lower than that obtained from the Paschen lines alone.  For
example, for GOODS-N 22384, $R$ normalized to $\ha$ results in a fit
to the Balmer lines that underpredicts the Pa5 (Pa$\beta$) line flux
by $\simeq 0.2$\,dex.  For the same galaxy, $R$ normalized to Pa5
yields a fit to the Paschen lines that underestimates the higher order
Balmer line fluxes (i.e., $\hg$ and higher) by $\ga 0.3$\,dex.
For these two galaxies, it is evident that the \citet{cardelli89} curve
cannot simultaneously reproduce all the line ratios with a single
value of $\ebmvneb$.  This apparent discrepancy can be resolved if the
dust optical depth is sufficiently large such that the emission
dominating the Balmer lines comes from regions of lower dust reddening
relative to the Paschen lines.  We return to this point below.




The bottom two panels show galaxies where the reddening derived from
the Balmer and Paschen lines are consistent with each other.  Finally,
as noted in Section~\ref{sec:sampleselection}, there are some galaxies
(not shown) where the Balmer lines indicate little or no reddening,
and which were excluded from the final sample used for the present
analysis.  Thus, the AURORA parent sample contains a mix of galaxies,
some where $\ebmvneb^{\rm B}< \ebmvneb^{\rm P}$, others where
$\ebmvneb^{\rm B}\simeq \ebmvneb^{\rm P}$, and some where $\ebmvneb
\simeq 0$.

\begin{figure}
  \epsscale{1.2}
  \includegraphics[width=1.0\linewidth]{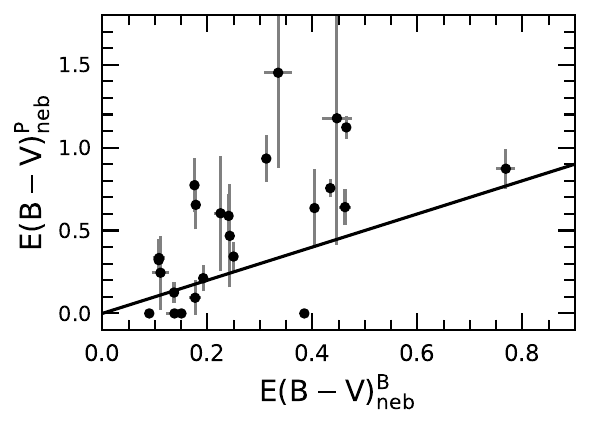}
    \caption{Comparison of $\ebmvneb^{\rm B}$ and $\ebmvneb^{\rm P}$
      for the 24 galaxies in the final sample.  The line of equality is
      indicated in black.}
    \label{fig:reddening}
\end{figure}

A comparison of $\ebmvneb^{\rm B}$ and $\ebmvneb^{\rm P}$ assuming the
Galactic extinction curve for all 24 galaxies in the final sample
(Figure~\ref{fig:reddening}) shows that 13 of them have $\ebmvneb^{\rm
  B} < \ebmvneb^{\rm P}$ at the $\le 1\sigma$ level.  The SMC
\citep{gordon03} extinction curve or the \citet{calzetti00}
attenuation curve results in $10$ and $11$ galaxies, respectively,
having $\ebmvneb^{\rm B} < \ebmvneb^{\rm P}$ at the $\le 1\sigma$
level.\footnote{As discussed in Section~\ref{sec:intro}, the SMC
    and Calzetti curves are commonly used to describe the reddening of
    the stellar continuum.  Here, we merely examine their effect on
    the derived nebular reddening to give a sense for the level of
    systematic variation in $\ebmvneb$ when assuming different curve
    shapes.}  These curves have a very similar shape to the Galactic
extinction curve at the wavelengths of the Balmer lines, resulting in
$\ebmvneb^{\rm B}$ that are on average 0.01 and 0.03\,mag bluer than
those derived with the Galactic extinction curve.  At the wavelengths
of the Paschen lines, the SMC and Calzetti curves are somewhat steeper
than the Galactic extinction curve, returning $\ebmvneb^{\rm P}$ that
are on average 0.10 and 0.17\,mag bluer than those derived with the
Galactic extinction curve.  In any case, regardless of which
extinction {\em or} attenuation curve is assumed, we find that roughly
half of the galaxies have nebular reddening derived from the Paschen
lines that is systematically larger than that derived from the Balmer
lines; i.e., $\ebmvneb^{\rm P}>\ebmvneb^{\rm B}$ (e.g., see also
\citealt{prescott22, gimenez22, reddy23a}).  Despite this
  systematic offset, there is a strong correlation between
  $\ebmvneb^{\rm P}$ and $\ebmvneb^{\rm B}$: a Spearman correlation
  test between these parameters yields a correlation coefficient of
  $\rho = 0.65$ and a probability of a null correletion of
  $p=5.7\times 10^{-4}$.  This correlation between Paschen and
  Balmer-derived reddening has been confirmed in previous work
  \citep{reddy23a, seille24}.

Formally, there are four galaxies where the Paschen lines alone
indicate negligible reddening, while the Balmer lines indicate
substantial reddening.  In these cases, the Paschen line ratios have
sufficiently large uncertainties and/or scatter about the expected
intrinsic values, and thus do not provide sufficient leverage to
independently constrain the reddening.  For all four galaxies, the
combination of any of the Paschen lines and one or more Balmer lines
indicates non-zero reddening.  Thus, regardless of the choice of
attenuation curve, we do not find significant evidence for galaxies
where the Balmer-derived reddening is {\em larger} than the reddening
derived using a combination of the Balmer and Paschen lines,
consistent with previous studies \citep{prescott22, reddy23a}.

\section{The Effective Attenuation Curve}
\label{sec:effcurve}

As noted in the previous section, the assumption of the Galactic or
SMC extinction curves, or the Calzetti attenuation curve, results in
discrepancies between the Balmer and Paschen-inferred reddening for at
least half the galaxies in the sample.  This result suggests that the
nebular dust attenuation curves deviate from the typically-assumed
Galactic extinction curve, and other common extinction and attenuation
curves \citep{forster01, prescott22, sanders25}.
The effective attenuation curve---or, alternatively, the arbitrarily
normalized effective attenuation curve, or ``relative'' attenuation
curve---can be calculated using the methodology first introduced in
\citet{reddy20}, and presented in Section~\ref{sec:effmeth}.
Section~\ref{sec:indandave} discusses the attenuation curves derived
for individual galaxies, as well as the average curve computed for the
sample.  The latter can be useful when applying dust corrections to
populations of galaxies.  The consistency of this average attenuation
curve with the recombination line ratios of galaxies outside the main
sample, including those at $z>3$, is explored in
Section~\ref{sec:highzcurve}.  Constraints on the normalization of the
nebular attenuation curve are described in Section~\ref{sec:rvnorm}.

\subsection{Methodology}
\label{sec:effmeth}

The attenuation in magnitudes at wavelength $\lambda$ can be expressed
in terms of the nebular reddening and the nebular dust attenuation
curve:
\begin{equation}
A_{\rm neb}(\lambda) = \ebmvneb k_{\rm neb}(\lambda).
\label{eq:amag}
\end{equation}
Equation~\ref{eq:eq2} can then be written in terms of $A_{\rm neb}(\lambda)$:
\begin{equation}
\frac{f(\lambda_1)}{f(\lambda_2)} = \frac{f_0(\lambda_1)}{f_0(\lambda_2)}\times 10^{-0.4[A_{\rm neb}(\lambda_1) - A_{\rm neb}(\lambda_2)]},
\label{eq:amag2}
\end{equation}
or
\begin{eqnarray}
A_{\rm neb}(\lambda_2) & = & 2.5\left[\log_{10}\left(\frac{f(\lambda_1)}{f(\lambda_2)}\right) - \log_{10}\left(\frac{f_0(\lambda_1)}{f_0(\lambda_2)}\right)\right] \nonumber \\
& & + A_{\rm neb}(\lambda_1).
\label{eq:amag3}
\end{eqnarray}
Following \citet{reddy20}, we can define a new quantity that depends only on the measured and intrinsic line ratios:
\begin{eqnarray}
A_{\rm neb}'(\lambda_2) & \equiv & A_{\rm neb}(\lambda_2) + [1-A_{\rm neb}(\lambda_1)] \nonumber \\
& = & 2.5\left[\log_{10}\left(\frac{f(\lambda_1)}{f(\lambda_2)}\right) - \log_{10}\left(\frac{f_0(\lambda_1)}{f_0(\lambda_2)}\right)\right] + 1. \nonumber \\
& & 
\label{eq:amag4}
\end{eqnarray}
Given Equation~\ref{eq:amag}, the relative attenuation curve can be written as
\begin{equation}
k_{\rm neb}'(\lambda) = \frac{A_{\rm neb}'(\lambda)}{A_{\rm neb}'(B) - A_{\rm neb}'(V)},
\label{eq:kprime4}
\end{equation}
where we have set $\lambda=\lambda_2$ and used the definition that
$\ebmvneb = A_{\rm neb}(B)-A_{\rm neb}(V)$.  For the remainder of
the analysis, $k_{\rm neb}'(\lambda)$ was shifted so that $k_{\rm
  neb}'(V) = 0$, thus implying that
\begin{eqnarray}
k_{\rm neb}(\lambda) = k_{\rm neb}'(\lambda) + \rv,
\label{eq:knebrv}
\end{eqnarray}
where $\rv$ is the value of the total attenuation curve at $V$-band
($5500$\,\AA).  The normalization constant, $\rv$, is discussed
further in Section~\ref{sec:rvnorm}.

$A_{\rm neb}'(\lambda)$ was calculated for each of the 24 galaxies in
the sample based on the observed and intrinsic line ratios using $\ha$
as the reference line ($\lambda_1$).  A fifth-order polynomial was
then fit to $A_{\rm neb}'(\lambda)$ versus $\mu = 1/\lambda$.  For
some objects in the sample, the noisier higher-order Balmer lines can
result in a non-monotonic polynomial fit to $A_{\rm neb}'(\lambda)$.
To enforce monotonicity, this polynomial was constrained to have a
derivative that is a perfect square (i.e., the square of a quadratic
function).  This requirement of monotonicity implies that the fit has
4 (rather than 6) degrees of freedom.  The fit was used to calculate
$A_{\rm neb}'(B)$ and $A_{\rm neb}'(V)$, where we adopt $4400$\,\AA\,
and $5500$\,\AA, respectively, for the effective wavelengths of the
$B$ and $V$ bands.  The $A_{\rm neb}'(\lambda)$ data points and
polynomial fit were then divided by $A_{\rm neb}'(B)-A_{\rm neb}'(V) =
\ebmvneb^{\rm eff}$ to calculate $k_{\rm neb}'(\lambda)$.  With the
previous definitions, $k_{\rm neb}'(\lambda)$ is equivalent to $k_{\rm
  neb}(\lambda)$ up to a normalization constant
(Equation~\ref{eq:knebrv}).

\begin{figure*}
  \epsscale{1.0}
  \includegraphics[width=1.0\linewidth]{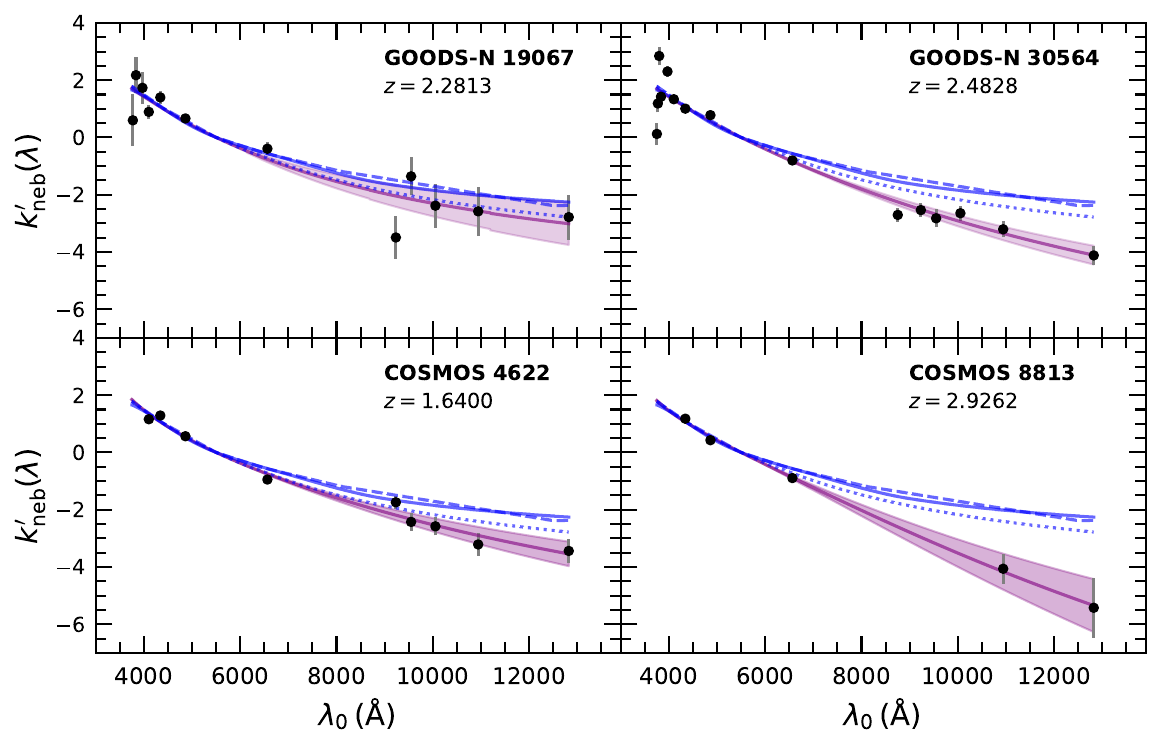}
    \caption{$k_{\rm neb}'(\lambda)$ versus $\lambda_0$ for four
      galaxies in the sample, with the best-fit fifth-order
      polynomials and $1\sigma$ confidence intervals shown by the
      purple lines and shaded regions, respectively.
Also shown in each panel are the Galactic, SMC, and Calzetti curves
(solid, dotted, and dashed blue lines, respectively), shifted so that
they pass through zero at $5500$\,\AA.}
    \label{fig:kquadprime_examples}
\end{figure*}

\subsection{Individual and Average Relative Nebular Attenuation Curves}
\label{sec:indandave}

Figure~\ref{fig:kquadprime_examples} shows $k_{\rm neb}'(\lambda)$ as
a function of wavelength for four representative galaxies in the
sample.  Also shown for comparison are the Galactic and SMC extinction
curves, and the Calzetti attenuation curve, all shifted so that their
value at $V$-band is zero.  The $k_{\rm neb}'(\lambda)$ for individual
galaxies show a diversity of shapes, from ones that are essentially
identical in shape to other commonly-adopted extinction and
attenuation curves (GOODSN-19067), to ones that deviate markedly
in shape from the common curves at $\lambda \ga 8000$\,\AA\,
(e.g., GOODSN-30564 and COSMOS-8813).

The inability to simultaneously reproduce all the Balmer and Paschen
line ratios with the commonly-adopted extinction or attenuation curves
for most of the galaxies in the sample, as discussed in
Section~\ref{sec:nebred}, suggests effective attenuation curves that
differ in shape from the commonly adopted ones.  This effect can be seen
directly in Figure~\ref{fig:kquadprime_examples}, where the curves of
three of the four galaxies shown (GOODSN-30564, COSMOS-4622, and
COSMOS-8813) have a steeper dependence on wavelength, particularly at
near-IR wavelengths, than the Galactic, SMC, and Calzetti curves.
This behavior cannot be attributed to inaccuracies in the relative
flux calibration between adjacent gratings.  Specifically, as noted in
Section~\ref{sec:slitloss}, all galaxies in the sample have
significantly detected continua and/or emission lines in the overlap
regions between gratings, allowing for accurate relative flux
calibration.  Additionally, 10 galaxies that show attenuation curves
that deviate from the Galactic extinction curve have one or more
Paschen lines in the same grating as $\ha$.  Finally, the
wavelength-dependent slitloss corrections inferred from the
transmission curves described in Section~\ref{sec:slitloss} result in
a relative shift between $k'(\lambda)$ at the wavelengths of $\ha$ and
Pa5 of $\Delta k'(\lambda) \simeq 0.27$ on average.  This shift is
considerably smaller than the typical deviation of $k'(\lambda)$ from
the value predicted by the Galactic extinction curve at the wavelength
of Pa5 (Figure~\ref{fig:kquadprime_examples}).  Therefore, the
deviations observed in the attenuation curves, when compared to other
standard extinction or attenuation curves, cannot be ascribed to flux
calibration uncertainties between gratings or uncertainties in
slitloss corrections. Instead, these deviations are likely to be of a
physical nature.

For the galaxies shown in Figure~\ref{fig:kquadprime_examples}, the
shapes of the curves at $\lambda \la 5500$\,\AA\, (spanning the
wavelength range of all the Balmer lines except $\ha$) is similar to
that of the Galactic extinction curve (or other common curves for that
matter), a result that could have been anticipated from the fact that
the Balmer line ratios alone can be well fit with the Galactic
extinction curve (e.g., Figure~\ref{fig:examples}).  It is for this
reason that $\ebmvneb^{\rm eff} \simeq \ebmvneb^{\rm B}$, since both
are determined at wavelengths proximate to the Balmer lines where the
shapes of the individual nebular attenuation curves are not
substantially different from that of the Galactic extinction curve.
The deviation in the shape of the nebular attenuation curve from other
standard curves only becomes apparent when considering the Balmer and
Paschen lines together, a point that is discussed further in
Section~\ref{sec:comparer20}.

\begin{figure}
  \epsscale{1.}
  \includegraphics[width=1.0\linewidth]{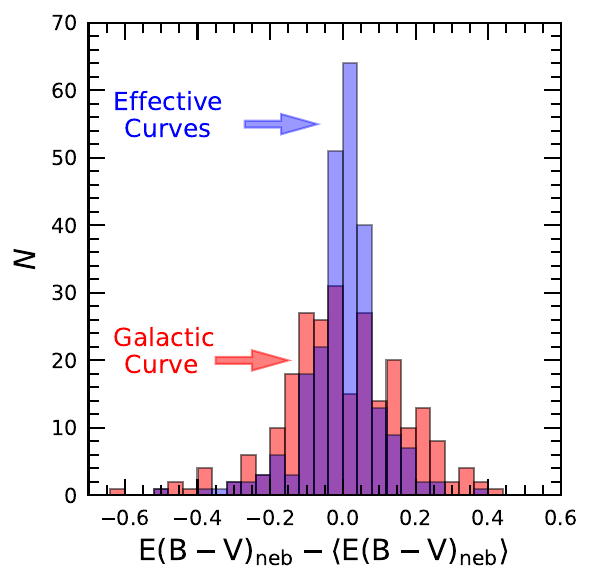}
    \caption{Difference in the reddening computed using multiple line
      ratios and the average reddening for individual galaxies, shown
      for all 24 galaxies in the sample, assuming the individual
      effective nebular attenuation curves (blue histogram) and the
      Galactic extinction curve (red histogram).}
   \label{fig:ebmvdiff}
\end{figure}

The reliability of the individual effective nebular attenuation curves
can be assessed by determining the consistency of the reddening values
derived from different line ratios for the same galaxy.  For each
galaxy, multiple estimates of $\ebmvneb$ were calculated using each
available line ratio relative to $\ha$.  The differences between these
multiple $\ebmvneb$ estimates and their average for all 24 galaxies,
are shown in Figure~\ref{fig:ebmvdiff}.  The red and blue histograms
indicate the distributions of these reddening differences, assuming
the Galactic extinction curve and the individual effective nebular
attenuation curves, respectively.  As expected, the individual
effective nebular attenuation curves yield a narrower spread in
$\ebmvneb$ calculated from multiple line ratios, compared to the
Galactic extinction curve.  The standard deviation of the reddening
differences computed with the individual nebular attenuation curves is
$\sigma = 0.10$, compared to $\sigma=0.17$ for the Galactic extinction
curve.  The difference in the spread of reddening values derived using
independent line ratios implies nebular attenuation curves that are
better able to simultaneously reproduce these line ratios relative to
the Galactic extinction curve.  Finally, we note that the
$\ebmvneb^{\rm eff}$ are on average $0.025$\,mag bluer when using line
fluxes corrected for stellar absorption based on the best-fit BPASS
CSF models (Section~\ref{sec:linemeas}).

\begin{figure}
  \epsscale{1.15}
  \includegraphics[width=1.0\linewidth]{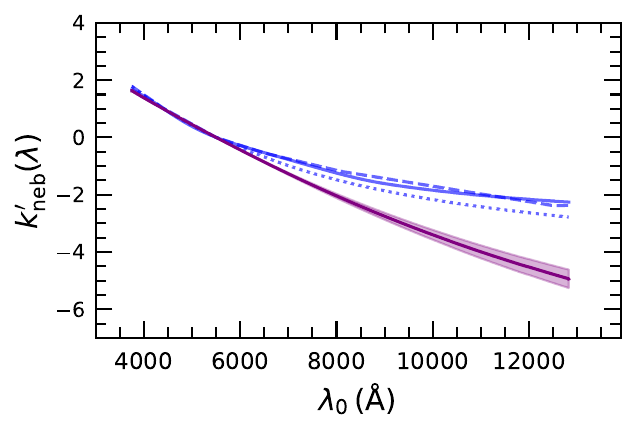}
    \caption{The average $k_{\rm neb}'(\lambda)$ curve, normalized so
      that $k_{\rm neb}'(V) = 0$, and its $1\sigma$ confidence
      interval are indicated by the solid purple and light shaded
      regions, respectively.  The curves expected for the Galactic,
      SMC, and Calzetti curves are shown by the solid, dotted, and
      dashed blue lines, respectively.}
   \label{fig:kquadprime_average}
\end{figure}

The average nebular dust attenuation curve, $\langle
k'(\lambda)\rangle$, was derived by simply averaging the $k'(\lambda)$
of the 24 galaxies, and is shown in
Figure~\ref{fig:kquadprime_average}.  The uncertainty in $\langle
k'(\lambda)\rangle$ includes both measurement errors and sample
variance.  The fifth-order polynomial fit to $\langle k_{\rm
  neb}'(\lambda)\rangle$ derived for the 24 galaxies is as follows:
\begin{eqnarray}
\langle k_{\rm neb}'(\lambda)\rangle & = & -14.198 + \frac{17.002}{\lambda/\mu{\rm m}} - \frac{8.086}{(\lambda/\mu{\rm m})^2} \nonumber \\
& & + \frac{2.177}{(\lambda/\mu{\rm m})^3} - \frac{0.319}{(\lambda/\mu{\rm m})^4} + \frac{0.021}{(\lambda/\mu{\rm m})^5},
\label{eq:effcurveshape}
\end{eqnarray}
normalized so that $k_{\rm neb}'(V)=0$ and valid over the range
$0.35\la \lambda \la 1.28$\,$\mu$m.  The average nebular dust
attenuation curve has slightly less curvature than that of common
extinction and attenuation curves at $\lambda \la 5500$\,\AA, and
exhibits a substantially steeper wavelength dependence than other
common curves at near-IR wavelengths, similar to the behavior noted
for the individual $k_{\rm neb}'(\lambda)$ curves shown in
Figure~\ref{fig:kquadprime_examples}.  Equation~\ref{eq:effcurveshape}
can be used to determine the dust-corrected ratio between two nebular
lines, as the ratio does not depend on the normalization of the
nebular attenuation curve.  However, we caution against using this
curve at wavelengths far beyond the range used to constrain it.
Absolute line fluxes do require some constraint or assumption on the
normalization of the nebular attenuation curve, a point that is
discussed further in Section~\ref{sec:rvnorm}.

\begin{figure}
  \epsscale{1.15}
  \includegraphics[width=1.0\linewidth]{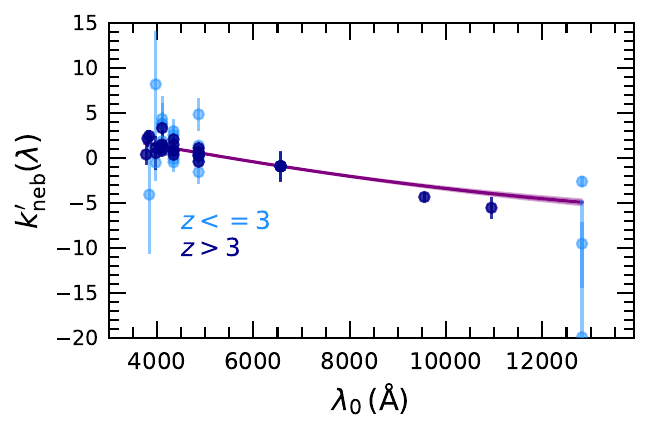}
    \caption{$k_{\rm neb}'(\lambda)$ versus $\lambda_0$ for 4 and 5
      galaxies at $z\le 3$ (light blue symbols) and $z>3$ (dark blue
      symbols), respectively, compared with the average curve found
      for the 24 galaxies in the main sample (purple curve).}
   \label{fig:kquadprime_highz}
\end{figure}

\subsection{Applicability to Higher-Redshift Galaxies}
\label{sec:highzcurve}

Though we have computed $\langle k_{\rm neb}'(\lambda)\rangle$ for a
subset of 24 galaxies in the AURORA sample, we can examine whether the
line ratios for galaxies with line coverage insufficient to make it
into the final sample are consistent with the average nebular
attenuation curve.  There are 9 additional galaxies with coverage of
at least 3 Balmer lines, 5 of which have redshifts $z>3$, for which we
were able to derive stringent constraints on $k_{\rm neb}'(\lambda)$.
The $k_{\rm neb}'(\lambda)$ for the 9 galaxies are shown in
Figure~\ref{fig:kquadprime_highz}, along with the average nebular
attenuation curve calculated using the main sample of 24 galaxies.
The individual $k_{\rm neb}'(\lambda)$ for the 9 additional galaxies,
and in particular those at $z>3$, scatter around the average curve,
suggesting no significant redshift evolution in the shape of the
curve.  Thus, our results tentatively suggest that the average curve
may be applied over a broad range of redshifts.  Larger samples of
galaxies at $z\ga 3$ with multiple detections of Balmer lines and
detections of higher-order Paschen lines (e.g., Pa6 and higher;
Figure~\ref{fig:redshifts}) will be needed to robustly test for any
possible redshift evolution in the shape of the nebular attenuation
curve, particularly when restricting such samples to similar ranges of
SFR, mass, metallicity, and other properties known to correlate with
dust attenuation.

\subsection{Normalization of the Nebular Dust Attenuation Curve}
\label{sec:rvnorm}

The $k_{\rm neb}'(\lambda)$ presented above is arbitrarily normalized
such that $k_{\rm neb}'(V) = 0$ (Section~\ref{sec:effmeth}).  Thus
$k_{\rm neb}(\lambda)$ at V band, i.e., $k_{\rm neb}(V)$, is simply
equivalent to $\rv$, the ratio of the total to selective extinction;
i.e., $\rv$ is directly related to the normalization of the dust
attenuation curve (Equation~\ref{eq:knebrv}).  Translating the
relative attenuation curve to a total effective attenuation curve is
typically accomplished by either extrapolating the relative curve to
some sufficiently long wavelength and forcing the total attenuation to
be zero at this wavelength (e.g., \citealt{calzetti97, battisti17,
  reddy15, reddy20}); or, in the case of a stellar reddening curve,
assuming that the integrated absorbed luminosity is re-radiated at
infrared wavelengths and using infrared luminosities to constrain the
normalization of the dust curve \citep{calzetti00}.  In principle,
this second method can be adapted to constrain the normalization of
the nebular attenuation curve by requiring that the inferred
dust-corrected nebular-based SFR agrees with the total bolometric SFR
taken as the sum of the unobscured UV-based SFR and the dust-obscured
IR-based SFR.  If there are reasons to suspect that such SFRs should
not agree---as is the case when one adopts an age-independent
conversion between luminosity and SFR for a galaxy undergoing a burst
of star formation where the $\hi$ recombination line flux responds
more quickly to changes in SFR than the UV and IR luminosities---then
this second method may not be accurate.

For simplicity, in the present analysis, we considered the first
method of constraining the normalization of the relative nebular dust
attenuation curve.  Similar to the behavior of other common
attenuation curves, the nebular attenuation curve is assumed to fall
to zero (``zero wavelength'') around $\lambda_0 \simeq 2.8$\,$\mu$m
(e.g., \citealt{calzetti97, gordon03, reddy15, battisti17};
c.f. \citealt{xue16, decleir22}).  In general, there are three primary
sources of uncertainty in the normalization of the curve: random error
from measurement uncertainties in $k_{\rm neb}'(\lambda)$, systematic
error from the functional form used to extrapolate the curve to long
wavelengths, and systematic error from the exact wavelength at which
the attenuation is forced to be zero.  Below, we consider two
functional forms of extrapolating the nebular dust attenuation curve
to long wavelength: a quadratic and linear function.  In
Section~\ref{sec:avecurvenorm}, we also consider a new approach to
constrain the normalization of the nebular dust attenuation curve.

We first considered a quadratic extrapolation of the attenuation curve
to long wavelengths.  Lower-upper (LU) decomposition was used to find
the quadratic function of $\mu = 1/\lambda$ whose value and derivative
(slope) match those of $k_{\rm neb}'(\lambda)$ at the longest
wavelength $\hi$ recombination line measured for each galaxy, and
where the quadratic function has a minimum value at $\lambda_0 =
2.8$\,$\mu$m.  This minimum value was subtracted from $k_{\rm
  neb}'(\lambda)$ for each galaxy to then arrive at the total
attenuation curve, $k_{\rm neb}(\lambda)$.  Likewise, LU decomposition
was used to find the linear function of $\mu = 1/\lambda$ whose value
and derivative (slope) match those of $k_{\rm neb}'(\lambda)$ at the
longest wavelength $\hi$ recombination line measured for each galaxy.
The value of this linear function at $\lambda_0 = 2.8$\,$\mu$m was
subtracted from $k_{\rm neb}'(\lambda)$ for each galaxy to then arrive
at the total attenuation curve, $k_{\rm neb}(\lambda)$.

For the quadratic extrapolation, the $\rv$ vary between $\rv \simeq
2.3$ to $14.2$, with a median value of $\rv = 5.5$ and a median
fractional uncertainty of $\sigma(\rv)/\rv = 0.22$.  The less freedom
offered by a simple linear-function extrapolation in $\mu$ results in
$\rv$ that are on average $\approx 25\%$ larger than those derived
from a quadratic extrapolation, varying in the range $\rv \simeq 2.5 -
19.8$ with a median value of $\rv = 6.8$ and a median fractional
uncertainty of $\sigma(\rv)/\rv = 0.24$.  In both extrapolations, we
find that the median $\rv$ is relatively large compared to the values
associated with standard dust extinction and attenuation curves.  We
return to this point in Section~\ref{sec:rvbackground}.

To determine the level of systematic uncertainty in $\rv$ stemming
from the choice of the zero wavelength, the $\rv$ values were
recalculated so that the attenuation curve falls to zero at
$3.4$\,$\mu$m.  In this case, the average and median $\rv$ are
approximately $4\%$ larger than the comparable values that assume a
zero wavelength of $2.8$\,$\mu$m.  The systematic uncertainty in $\rv$
due to the choice of zero wavelength is appreciably smaller than
either the measurement uncertainty or the systematic uncertainty
arising from the function used to extrapolate the curve to long
wavelengths.

There is also a small systematic uncertainty in $\rv$ arising from the
choice of SPS model used to correct the $\hi$ line fluxes for stellar
absorption.  As noted in Section~\ref{sec:linemeas}, the stellar
absorption corrections are based on the best-fitting \citet{conroy09}
flexible delayed-$\tau$ models.  Adopting the simpler BPASS CSF models
results in derived $\rv$ that are on average $\Delta\rv \simeq 1.31$
larger than the default values.

\section{Discussion}
\label{sec:discussion}

A key result of this analysis is the finding of a relatively high
median $\rv$ and a large spread in $\rv$ (Section~\ref{sec:rvnorm})
compared to standard extinction and attenuation curves, including the
Galactic, SMC, and Calzetti curves.  Section~\ref{sec:rvbackground}
summarizes the relationship between $\rv$ and the slope and
normalization of the dust attenuation curve, and the $\rv$ that are
typically associated with changes in dust grain composition and
geometry.  Section~\ref{sec:subunity} presents a model that
encapsulates geometrical variations in the dust distribution by the
dust covering fraction.  The final normalization of the nebular dust
attenuation curves and total nebular attenuation curve are presented
in Section~\ref{sec:avecurvenorm}.  A comparison of the attenuation
curve found in this analysis and that of \citet{reddy20} is discussed
in Section~\ref{sec:comparer20}.  Finally, Section~\ref{sec:rvdiscuss}
presents the physical interpretation of the normalization of the dust
curve.

\subsection{Dependence of $\rv$ on Dust Grain Composition and Geometry}
\label{sec:rvbackground}

The total-to-selective extinction ratio, $\rv$, is related to the
optical ``slope'' of the nebular attenuation curve, commonly defined
as $A_{\rm neb}(B)/A_{\rm neb}(V)$, by the following equation:
\begin{equation}
\frac{A_{\rm neb}(B)}{A_{\rm neb}(V)} = \frac{1}{\rv} + 1.
\label{eq:defrv}
\end{equation}
The optical slope can also be expressed as $k_{\rm B}/k_{\rm V}$.  By
definition, $k_{\rm B}-k_{\rm V} \equiv 1$.  However, the ratio,
$k_{\rm B}/k_{\rm V}$, can be arbitrarily large, with a minimum value
of approximately unity for an almost flat attenuation curve where $\rv
\rightarrow \infty$.\footnote{Since $k_{\rm B}-k_{\rm V} \equiv 1$,
  $k_{\rm B}/k_{\rm V} \approx 1$ implies that $k_{\rm B}\gg 1$ and
  $k_{\rm V}\gg 1$, and $\rv \gg 1$.}  Rearranging
Equation~\ref{eq:defrv} and combining with Equation~\ref{eq:amag}
yields:
\begin{equation}
\rv = \frac{A_{\rm neb}(V)}{A_{\rm neb}(B)-A_{\rm neb}(V)} = \frac{\ebmvneb k_{\rm neb}(V)}{A_{\rm neb}(B)-A_{\rm neb}(V)}.
\end{equation}
Since $\ebmvneb = A_{\rm neb}(B) - A_{\rm neb}(V)$, the above equation reduces to
\begin{equation}
\rv = k_{\rm neb}(V) = k_{\rm neb}'(V)+C,
\end{equation}
where $C$ is the constant of normalization that relates $k_{\rm
  neb}'(\lambda)$ and $k_{\rm neb}(\lambda)$.  Since $k_{\rm
  neb}'(V)\equiv 0$ (Equation~\ref{eq:knebrv}), the above equation
simply says that $\rv$ is a constant providing the offset between
$k_{\rm neb}'(\lambda)$ and $k_{\rm neb}(\lambda)$.  Thus, $\rv$ is
related to both the slope and normalization of the nebular attenuation
curve, and is equivalent to the value of the total nebular attenuation
curve at $V$-band.  As the normalization of the nebular attenuation
curve increases, $\rv$ increases, and the optical slope of the curve
becomes shallower.

In the context of line-of-sight {\em extinction} curves, variations in
$\rv$ have been linked to differences in the dust grain size
distribution \citep{fitzpatrick99, draine03, li21}.  Environments
where small dust grains are more abundant---such as those affected by
supernovae shocks, which thermally sputter large grains, or regions
experiencing grain shattering in the warm ionized ISM (e.g.,
\citealt{hirashita10})---typically exhibit steeper extinction curves.
Such curves result in more substantial extinction at UV and optical
wavelengths, leading to a lower $\rv$.  Conversely, environments that
favor the growth of large dust grains---such as through the
coagulation of smaller grains in the ISM (e.g., \citealt{hirashita12,
  galliano18})---tend to produce flatter, or ``grayer,'' extinction
curves, corresponding to higher $\rv$.  The balance between dust
destruction and grain growth determines the overall grain-size
distribution and, by extension, the slope and normalization of the
extinction curve.  The composition of the grains (silicates versus
graphitic) may also affect the shape of the extinction curve
\citep{li21}.  Observed sightlines in the Milky Way as well as towards
the SMC and LMC exhibit varying extinction curves with $\rv$ values
ranging from 2.7 to 5.5 (e.g., \citealt{cardelli89, fitzpatrick99,
  clayton00, gordon03, fitzpatrick09}).

\begin{figure}
  \epsscale{1.2}
  \includegraphics[width=1.0\linewidth]{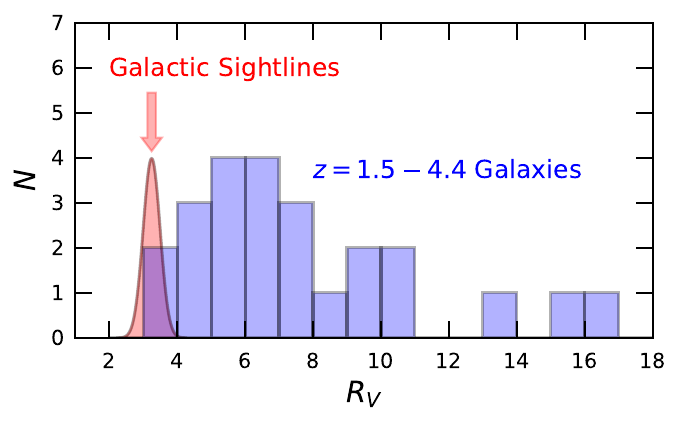}
    \caption{Histogram of $\rv$ for the 24 galaxies in the sample
      (shown in blue).  These effective $\rv$ values were obtained
      using the methodology explained in
      Section~\ref{sec:avecurvenorm}.  For comparison, the arbitrarily
      normalized distribution of $\rv$ derived by \citet{zhang23} for
      3 million stars from the Large Sky Area Multi-Object Fiber
      Spectroscopy Telescope (LAMOST) survey is shown in red.  These
      authors find a Gaussian distribution of $\rv$ with a mean of
      3.25 and standard deviation of 0.25.}
    \label{fig:rvcomphist}
\end{figure}

In the case of {\em attenuation} curves, where scattering of light
into the line of sight, nonuniform column density distributions, or
spatial variations in optical depth are significant, the value of
$\rv$ can be arbitrarily large \citep{narayanan18, trayford20}.  The
median $\rv$ values inferred for the AURORA sample range from $\simeq
5.5$ to $6.8$ depending on whether a quadratic or linear extrapolation
to long wavelength is used (Section~\ref{sec:rvnorm}), with some
galaxies exhibiting $\rv \ga 10$.  These values are notably higher
than those typically observed along Galactic sightlines, which usually
have $\rv \la 5.5$ across a variety of environments, as illustrated in
Figure~\ref{fig:rvcomphist}.  Consequently, the large $\rv$ inferred
for the nebular attenuation curves suggest that their variation is
driven primarily by the geometry of the dust relative to the OB
associations, rather than by differences in the dust grain size
distribution.  Specifically, the light from OB associations along low
dust-column-density (and thus relatively unobscured) sightlines
reduces the effective nebular attenuation at short wavelengths.  This
leads to a flattening of the attenuation curve and, consequently, a
larger $\rv$.

While we cannot definitely rule out the possibility that some of the
variation in the nebular dust attenuation curves is due to differences
in dust grain composition or size distribution, there are several
pieces of evidence that support a scenario of a non-uniform dust
distribution towards OB associations.  First, the very high $\rv\ga 5$
values found for most galaxies in the sample relative to the values
found along different environments in the Milky Way
(Figure~\ref{fig:rvcomphist}) suggest the importance of the dust-stars
geometry.  Second, prior evidence of non-unity gas and metal covering
fractions in similar galaxies from other studies (e.g.,
\citealt{shapley03, heckman11, berry12, jones13, alexandroff15,
  trainor15, henry15, vasei16, reddy16b, steidel18, du18, gazagnes18,
  trainor19, jaskot16, harikane20, reddy22}, among many others) imply
a non-unity covering fraction of dust as well.  Third, the
near-ubiquitous presence of very massive O-star P-Cygni photospheric
lines (e.g., $\civ$\,$\lambda\lambda 1548,1550$ for $\ga
30$\,$M_\odot$ O stars and $\siiv$\,$\lambda\lambda 1393, 1402$ in O
supergiants) and Wolf Rayet features (e.g., $\heii$\,$\lambda 1640$)
in the far-UV spectra of high-redshift star-forming galaxies (e.g.,
\citealt{pettini00, shapley03}) implies that some fraction of even the
youngest OB associations must lie along relatively unobscured
sightlines.  Fourth, spatial offsets between UV and dust emission
(e.g., \citep{willott15, pentericci16, schouws22, bowler22, inami22})
imply some fraction of the OB associations are in dust-free regions.
Fifth, differences in the reddening of Balmer and Paschen lines from
integrated line measurements (e.g., \citealt{reddy23a}) and, most
recently, spatial offsets between $\ha$ and Paschen emission
\citep{lorenz25} support a scenario of non-uniform dust distribution
towards OB associations.


\subsection{Modeling with a Sub-unity Covering Fraction of Dust}
\label{sec:subunity}

The high $\rv$ inferred for the sample, and the association of such
high values to variations in the dust-stars geometry (e.g.,
\citealt{narayanan18, trayford20}), suggest that a uniform foreground
screen (i.e., with a $100\%$ covering fraction of dust) may not be an
appropriate description for at least half the galaxies in the sample.
In Section~\ref{sec:nebred}, we suggested that differences in the
reddenings derived from the Balmer and Paschen lines may indicate the
presence of regions with high dust column densities, causing the
shorter-wavelength recombination lines which are more heavily
attenuated by dust to be dominated by emission from the unreddened
regions of the galaxy (see also \citealt{reddy23a}).  Along these
lines, \citet{lorenz25} use JWST/NIRCam medium-band photometry to
construct spatially-resolved $\ha$ and Pa$\beta$ maps for a sample of
14 star-forming galaxies at $1.3\le z\le 2.4$.  They find that $\la
0\farcs 1$ offsets between $\ha$ and Pa$\beta$ emission are common,
suggesting the presence of heavily-reddened star formation.  The
implied non-uniform distribution of dust along the sightlines to OB
associations in galaxies has been invoked to explain other findings,
including the offset between the reddening of the nebular regions and
stellar continuum in high-redshift galaxies \citep{reddy20, lorenz23,
  lorenz24}, and may also account for much of the scatter between dust
attenuation, as quantified by the ratio of the infrared-to-UV
luminosity, and UV spectral slope (i.e., the IRX-$\beta$ diagram;
\citealt{burgarella05, reddy06a, bouwens16a, buat18, reddy18a,
  salim19}; see also review by \citealt{salim20}).  In principle, the
Calzetti curve includes the effects of a non-unity covering fraction
of dust on the attenuation of the stellar continuum of nearby
starburst galaxies \citep{calzetti00}, but even this curve is unable
to reconcile the full suite of Balmer and Paschen emission-line ratios
accessible in this study for a majority of galaxies in the present
sample.  The primary distinction here, of course, is that we are
concerned with the distribution of dust along the sightlines to the
$\hii$ regions around massive stars, since that is where the
recombination lines originate, rather than the distribution of dust
towards the stars dominating the (non-ionizing) stellar continuum.
Here, we discuss a model of the attenuation that allows for a
non-uniform distribution of dust.

\subsubsection{Formalism}
\label{sec:subunityformalism}

\begin{figure}
  \epsscale{1.2}
  \includegraphics[width=1.0\linewidth]{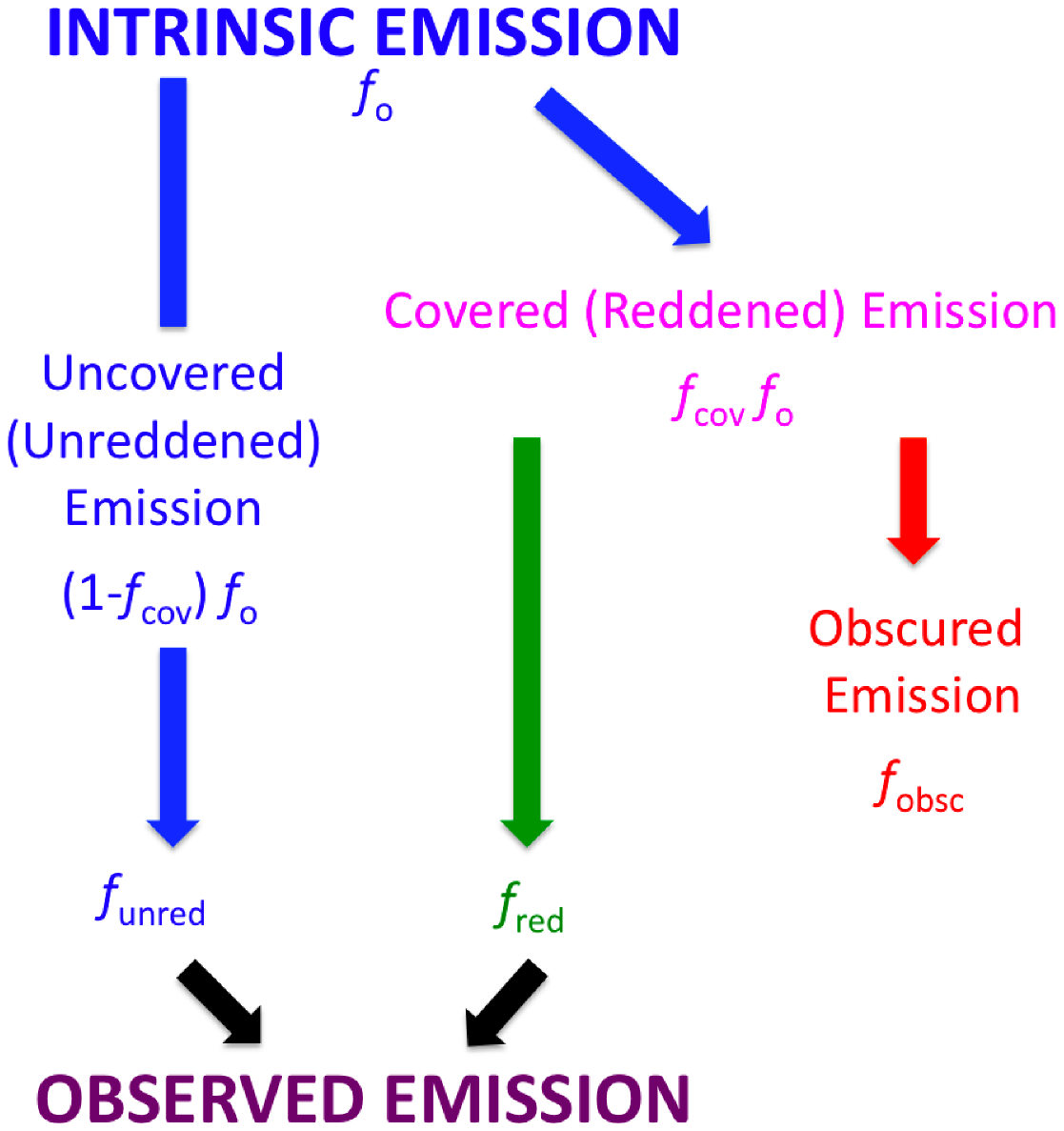}
    \caption{Schematic diagram of the covering-fraction model.  Some
      fraction of the $\hi$ line emission escapes the galaxy unimpeded
      by dust, denoted as the uncovered (unreddened) emission (blue).
      The remaining fraction is reddened by dust (covered reddened
      emission), some of which exits the galaxy (green) and some of
      which is absorbed by dust and reemitted in the infrared
      (obscured emission, red).  The uncovered and reddened emission
      both contribute to the total observed $\hi$ line emission.}
    \label{fig:schematic}
\end{figure}

Variations in dust optical depth within a galaxy can be parameterized
by introducing another factor in addition to the reddening, namely the
dust covering fraction, $\fcov$.  In this section, we build upon the
simple foreground screen model expressed in Equation~\ref{eq:eq1} to
account for a non-unity covering fraction of dust, akin to the
formalism used to model the rest-frame UV continuum light of galaxies
with a non-unity covering fraction of gas and dust (e.g.,
\citealt{reddy16a, reddy16b, steidel18}; see also
\citealt{forster01, prescott22}).  We assume a simple model
(referred to as the ``covering-fraction model'') in which some
fraction of the intrinsic line emission escapes unimpeded, while the
remaining fraction is subject to dust reddening, as shown in the
schematic diagram of Figure~\ref{fig:schematic}.  We define the uncovered,
or unreddened, emission as 
\begin{eqnarray}
f_{\rm unred}(\lambda) & = & (1-\fcov)f_0(\lambda)
\end{eqnarray}
and the covered, or reddened, emission as 
\begin{eqnarray}
f_{\rm red}(\lambda) & = & \fcov f_0(\lambda) \times 10^{-0.4\ebmvneb^{\rm cov}k_{\rm neb}(\lambda)},
\end{eqnarray}
where $\fcov$ denotes the covering fraction of dust, and $\ebmvneb^{\rm cov}$ is the reddening of the  dust-covered portion of the line emission.
In this case, the sum
of the unreddened and reddened line emission constitutes the observed
line flux:
\begin{eqnarray}
f(\lambda) & = & f_{\rm unred}(\lambda) + f_{\rm red}(\lambda) \nonumber \\
           & = & (1-\fcov)f_{0}(\lambda) + \nonumber \\
& & \fcov f_{0}(\lambda)\times 10^{-0.4\ebmvneb^{\rm cov}k_{\rm neb}(\lambda)}.
\label{eq:subunityeq1}
\end{eqnarray}
The first and second terms of this equation correspond to the
unreddened and reddened components, respectively. 
In this model, a
fixed fraction of the {\em intrinsic} flux is subject to reddening
($\fcov$) or escapes unimpeded ($1-\fcov$), independent of wavelength.
The actual {\em fraction} of the intrinsic flux extinguished by dust (i.e.,
the ratio of the ``obscured emission'' to the intrinsic emission as
depicted in Figure~\ref{fig:schematic}) is, of course, wavelength
dependent, and is given by
\begin{eqnarray}
{\mathcal{F}}_{\rm obsc}(\lambda) & = & \frac{f_{\rm obsc}(\lambda)}{f_0(\lambda)} \nonumber \\
& = & \frac{\fcov f_0(\lambda) [1-10^{-0.4\ebmvneb^{\rm cov}k_{\rm neb}(\lambda)}]}{f_0(\lambda)} \nonumber \\
& = & \fcov [ 1 - 10^{-0.4\ebmvneb^{\rm cov} k_{\rm neb}(\lambda)}].
\label{eq:obscuredfrac}
\end{eqnarray}

\begin{figure}
  \epsscale{1.2}
  \includegraphics[width=1.0\linewidth]{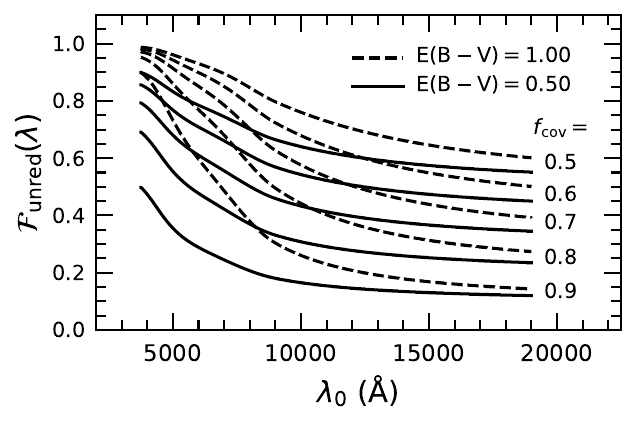}
    \caption{The fraction of observed line emission coming from the
      unobscured regions as a function of wavelength assuming a
      sub-unity covering fraction of dust and the Galactic extinction
      curve, as described by Equation~\ref{eq:subunityeq1}.  The solid
      and dashed curves correspond to $\ebmvneb^{\rm cov}=0.50$ and
      $1.00$ respectively, for the different values of $\fcov$
      indicated in the figure.  For a fixed covering fraction of dust,
      a smaller fraction of the observed emission arises from the
      unreddened regions for the longer-wavelength Paschen lines
      relative to the shorter-wavelength Balmer lines.}
    \label{fig:unredfrac}
\end{figure}

The fraction of the {\em observed} line emission that emerges from the
unreddened regions is also wavelength dependent.  For a
given combination of $\fcov$ and $\ebmvneb^{\rm cov}$, the fraction of
light emerging from the unreddened regions, $\mathcal{F}_{\rm
  unred}(\lambda)$---defined as the ratio of the uncovered
(unreddened) emission to the total observed emission, referring again to
Figure~\ref{fig:schematic}--- decreases with increasing wavelength.
Mathematically, 
\begin{eqnarray}
\mathcal{F}_{\rm unred}(\lambda) & = & \frac{f_{\rm unred}(\lambda)}{f(\lambda)} \nonumber \\
&  = & \frac{f_{\rm unred}(\lambda)}{f_{\rm unred}(\lambda)+f_{\rm red}(\lambda)} \nonumber \\
& = & \frac{1-\fcov}{1-\fcov + \fcov \times 10^{-0.4\ebmvneb^{\rm cov}k_{\rm neb}(\lambda)}}.
\label{eq:unredfrac}
\end{eqnarray}
These fractions are shown as a function of wavelength in
Figure~\ref{fig:unredfrac}, for two different combinations of $\fcov$
and $\ebmvneb^{\rm cov}$ assuming a ``line-of-sight'' extinction given
by the Galactic extinction curve.  For example, for $\fcov = 0.70$ and
$\ebmvneb^{\rm cov} = 0.50$, the fraction of the observed $\ha$
emission emerging from low-dust-optical-depth (i.e., unreddened)
regions is $\mathcal{F}_{\rm unred}=0.58$.  For the same conditions, $\mathcal{F}_{\rm
  unred}=0.39$ at the wavelength of Pa5 ; i.e., a larger fraction of
the observed Pa5 flux comes from the reddened regions compared to
$\ha$.  For the same $\fcov$ and $\ebmvneb^{\rm cov} = 1.00$, the
unreddened percentage increases to $82\%$ for $\ha$ and $48\%$ for
Pa5; i.e., at a fixed $\fcov$, an increase in $\ebmvneb^{\rm cov}$
leads to a decrease in the fraction of observed flux coming from the
reddened regions and an increase in the fraction from unreddened
regions, for any given emission line.

It is important to emphasize that $\mathcal{F}_{\rm unred}(\lambda)$
represents the fraction of the {\em observed}, rather than the
intrinsic, light that is unimpeded by dust.  Accordingly,
$\mathcal{F}_{\rm unred}(\lambda)$ may be large under two conditions:
(a) when the covering fraction is small (for $\fcov=0$,
$\mathcal{F}_{\rm unred}=1$ at all wavelengths), or (b) when
$\ebmvneb^{\rm cov}$ is large.  In the latter case, the fraction of
the observed line emission from regions with high $\ebmvneb^{\rm
  cov}$---i.e., the heavily-reddened regions---decreases, leading to a
higher value of $\mathcal{F}_{\rm unred}(\lambda)$ at a given $\fcov$.
If $\ebmvneb^{\rm cov}$ is extremely large (i.e., $\ebmvneb^{\rm
  cov}\gg 1$) such that even the Paschen lines become optically-thick,
then $\mathcal{F}_{\rm unred}$ will be uniformly high for all the
Balmer and Paschen lines.  In such a scenario, mid- or far-IR line
and/or dust continuum measurements may be needed to determine whether
the covering fraction is high or low.

\subsubsection{Distribution of $\ebmvneb^{\rm cov}$ and $\fcov$}
\label{sec:covfracparms}

\begin{figure}
  \epsscale{1.2}
  \includegraphics[width=1.0\linewidth]{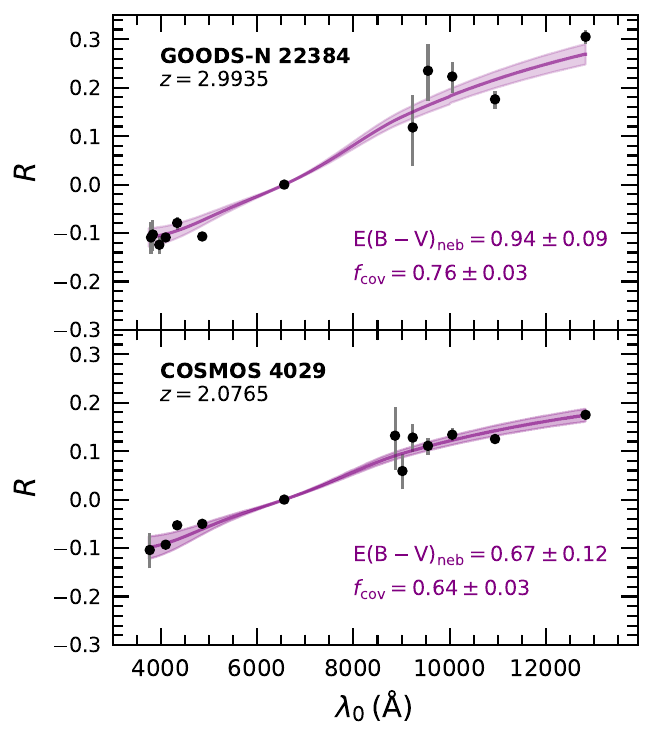}
    \caption{$R$ versus rest-frame wavelength for two galaxies in the
      sample.  The measured values and uncertainties in $R$ are
      indicated by the black points and errorbars.  The solid lines
      and shaded regions indicate the best-fit models that incorporate
      a sub-unity covering fraction of dust and $68\%$ confidence
      intervals of those fits, respectively.  The best-fit values of
      $\fcov$ and $\ebmvneb^{\rm cov}$ for the Galactic extinction
      curve are indicated in each panel.}
    \label{fig:subunity_examples}
\end{figure}

In the covering-fraction model, $R$---as defined in Equation~\ref{eq:eq3}---can
be expressed as follows:
\begin{equation}
R = \log_{10}\left[\frac{1-\fcov+\fcov\times 10^{-0.4\ebmvneb^{\rm cov}k_{\rm neb}(\lambda_1)}}{1-\fcov+\fcov\times 10^{-0.4\ebmvneb^{\rm cov}k_{\rm neb}(\lambda_2)}}\right].
\label{eq:eq6}
\end{equation}
Again, $\ha$ is used as the reference line.  Fitting the $R$ measured
for each galaxy using Equation~\ref{eq:eq6} yields an estimate of the
best-fit $\fcov$ and $\ebmvneb^{\rm cov}$.  The covering-fraction model fits
with the Galactic extinction curve for $k_{\rm neb}$ are shown in
Figure~\ref{fig:subunity_examples} for two galaxies in the sample
having $\ebmvneb^{\rm B}< \ebmvneb^{\rm P}$.  While the Galactic
extinction curve with $\fcov = 1$ is not able to simultaneously
predict all the Balmer and Paschen line ratios
(Section~\ref{sec:nebred}), a model with $\fcov < 1$ is able to
reproduce all the line ratios for these two galaxies and, in general,
the full sample of 24 galaxies.  Adopting an SMC or Calzetti curve for
$k_{\rm neb}$ results in $\ebmvneb^{\rm cov}$ that are on average
$0.02$\,mag redder and $0.13$\,mag bluer, respectively, than those
obtained with the Galactic extinction curve.  Likewise, relative to
the covering fractions derived with the Galactic extinction curve,
$\delta\fcov = 0.03$ smaller and $0.03$ larger, respectively, for the
SMC and Calzetti curves.  Regardless of the adopted extinction or
attenuation curve, the covering-fraction model provides improved fits for $R$
versus $\lambda_0$ relative to the model which assumes a unity covering
fraction.

\begin{figure}
  \epsscale{1.1}
  \includegraphics[width=1.0\linewidth]{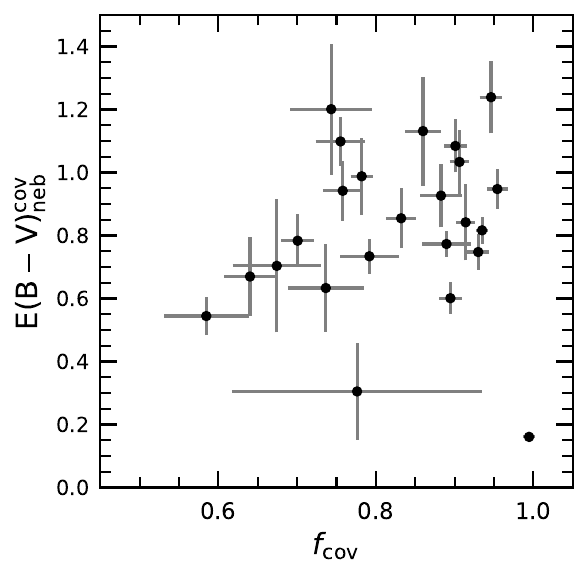}
    \caption{Distribution of $\fcov$ and $\ebmvneb^{\rm cov}$ for the
      full sample of galaxies, derived from the model with a sub-unity
      dust covering fraction.}
    \label{fig:subunity_parms}
\end{figure}

The distribution of $\fcov$ and $\ebmvneb^{\rm cov}$ for the full
sample is displayed in Figure~\ref{fig:subunity_parms}.  Covering
fractions vary from $\simeq 0.50$ to unity.  Note that there are
marginal anticorrelations between the random uncertainties in $\fcov$
and $\ebmvneb^{\rm cov}$ on the one hand, and $\fcov$ on the other;
i.e., the random uncertainties in $\fcov$ and $\ebmvneb^{\rm cov}$ are
on average larger at lower $\fcov$.  The larger uncertainties for
these galaxies may be due to lower $S/N$ in the lines, fewer number of
lines, or a smaller wavelength baseline of the lines being fit, which
may indicate that the covering-fraction modeling prefers lower $\fcov$
for such galaxies.  In Appendix~\ref{sec:covfracbias}, we present a
discussion of these possibilities, and the results of a simulation to
test for biases in the derived $\fcov$ and $\ebmvneb^{\rm cov}$.  From
this analysis, we conclude that the larger uncertainties for galaxies
with lower $\fcov$ is primarily due to lower $S/N$ in the lines being
fit, and that there is no biasing effect that would cause such
galaxies to be preferentially fit with lower $\fcov$.

The improved fits to $R$ provided by the covering-fraction model are
perhaps not surprising, given that this model introduces a second free
parameter---the covering fraction.  However, it offers a more accurate
representation of a non-uniform distribution of dust, which has been
suggested by previous studies, and is also supported by the high $\rv$
inferred for the nebular dust attenuation curves
(Section~\ref{sec:rvnorm}).

Finally, we note that corrections for stellar absorption based on the
BPASS CSF models result in inferred $\ebmvneb^{\rm cov}$ that are on
average 0.31\,mag redder than the default values.  Similarly, assuming
the BPASS CSF models to correct for stellar absorption results in
$\fcov$ that are on average 0.03 higher than the default values.

\subsubsection{Light-weighted Reddening}
\label{sec:lwred}

For the covering-fraction model, the fraction of the observed line
luminosity that emerges from the unreddened regions of a galaxy is
wavelength dependent, as noted earlier (Figure~\ref{fig:unredfrac}).
Thus, the reddening that one measures for a galaxy will depend on the
wavelength of the lines used to measure the reddening: i.e., the
light-weighted reddening, $\ebmvneb^{\rm lw}$, also varies with
wavelength.  Specifically, the reddening computed from the (shorter
wavelength) Balmer lines alone will be weighted more heavily towards
the unreddened sightlines, while the reddening computed from the
(longer wavelength) Paschen lines alone will be weighted towards the
reddened sightlines.  This variation naturally occurs because the
shorter wavelength lines are more heavily attenuated by dust than the
longer wavelength lines for a given dust column density (or
reddening).  The consistency of the covering-fraction model can be
checked by comparing the light-weighted reddening at a given
wavelength to the reddening deduced using only the recombination lines
spanning a similar wavelength range.  Any discrepancies between the
two can provide insight into the accuracy of the model assumptions and
the distribution of dust and stars in the galaxy.

The light-weighted reddening, $\ebmvneb^{\rm lw}$, can be expressed as
the product of two factors: the reddening of the covered portion of
light ($\ebmvneb^{\rm cov}$) and the fraction of the observed line
luminosity that is reddened, $\mathcal{F}_{\rm red}(\lambda) = 1-\mathcal{F}_{\rm unred}(\lambda)$, which
can be obtained from Equation~\ref{eq:unredfrac}, i.e.,
\begin{eqnarray}
\ebmvneb^{\rm lw} & = & \ebmvneb^{\rm cov} \times \mathcal{F}_{\rm red}(\lambda) \nonumber \\
                 & = & \ebmvneb^{\rm cov} \times \nonumber \\
& & \left[\frac{\fcov\times 10^{-0.4\ebmvneb^{\rm cov}k_{\rm neb}(\lambda)}}{1-\fcov + \fcov\times 10^{-0.4\ebmvneb^{\rm cov}k_{\rm neb}(\lambda)}}\right].
\label{eq:lwred}
\end{eqnarray}
Since $\mathcal{F}_{\rm red}$ is larger at longer wavelengths for a given
$\fcov$ and $\ebmvneb^{\rm cov}$ (see Figure~\ref{fig:unredfrac}), the
reddening measured at these wavelengths will be weighted more heavily
towards the reddened regions, leading to a higher value of the
light-weighted reddening, $\ebmvneb^{\rm lw}$, in comparison to the
shorter-wavelength Balmer lines.

\begin{figure}
  \epsscale{1.1}
  \includegraphics[width=1.0\linewidth]{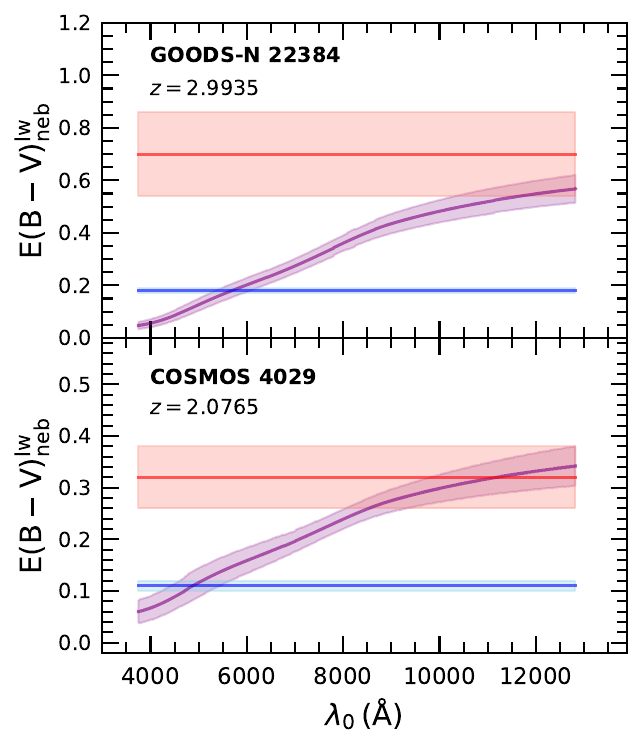}
    \caption{Light-weighted reddening, $\ebmvneb^{\rm lw}$, and its
      $1\sigma$ confidence interval as a function of wavelength
      (purple curve and shaded regions, respectively) for GOODSN-22384
      (top panel) and COSMOS-4029 (bottom panel).  For comparison,
      $\ebmvneb^{\rm B,P}$ and their $1\sigma$ uncertainties are
      indicated by the horizontal blue and red lines and shaded
      regions, respectively.}
    \label{fig:ebmvfobs_examples}
\end{figure}

The light-weighted reddening as a function of wavelength for two
galaxies in the sample (GOODSN-22384 and COSMOS-4029) is shown in
Figure~\ref{fig:ebmvfobs_examples}.  For comparison, $\ebmvneb^{\rm
  B}$, $\ebmvneb^{\rm P}$, and their associated uncertainties are also
show in this figure.  For these two galaxies, $\ebmvneb^{\rm lw}$
averaged over the wavelength ranges of the Balmer and Paschen lines is
similar to the reddening computed from either the Balmer lines alone,
or the Paschen lines alone, assuming a uniform screen of
dust.  More generally, we find that the covering-fraction model
predicts light-weighted reddenings that are consistent with those
deduced from either the Balmer or Paschen lines for most galaxies in
the sample.

\begin{figure}
  \epsscale{1.1}
  \includegraphics[width=1.0\linewidth]{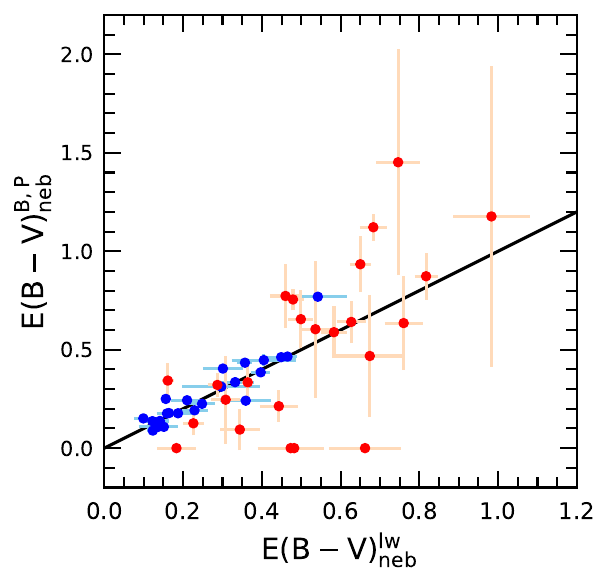}
    \caption{Comparison of the light-weighted reddening from the
      covering-fraction model computed at $5400$ and $9500$\,\AA\, with
      $\ebmvneb^{\rm B}$ (blue points) and $\ebmvneb^{\rm P}$ (red
      points).  The line of equality is indicated in black.}
    \label{fig:ebmvlw}
\end{figure}

To illustrate this, Figure~\ref{fig:ebmvlw} compares $\ebmvneb^{\rm
  lw}$ calculated from the covering-fraction modeling with the
$\ebmvneb$ computed from either the Balmer (blue points) or Paschen
lines (red points) assuming unity covering fraction of dust for the
full sample of galaxies.  The light-weighted reddening is computed at
$5400$ and $9500$\,\AA, roughly corresponding to the mean wavelengths
of the Balmer and Paschen lines, respectively, available for most
galaxies.  As noted in Section~\ref{sec:nebred}, the Balmer lines
provide more constraining power on $\ebmvneb$ because of their
sensitivity to the reddening and their typically higher $S/N$ than the
Paschen lines.  For the tightly constrained reddening deduced from the
Balmer lines alone, we find that the covering-fraction model
successfully predicts $\ebmvneb^{\rm B}$ for all but one of the
galaxies in the sample.  This one exception is COSMOS-4622, where
$\ebmvneb^{\rm lw} = 0.542 \pm 0.075$ computed at $\sim 5400$\,\AA,
while $\ebmvneb^{\rm B} = 0.769\pm 0.018$.  This discrepancy is due to
the fact that the line ratios for this particular galaxy are not as
well fit by the covering-fraction model, as evidenced by a somewhat
lower $\chi^2$ value for the fit relative to to the $\chi^2$ values
for most other galaxies in the sample.

The most notable outliers in the comparison of the light-weighted
reddening at $9500$\,\AA\, and the Paschen-derived reddening are the
same four galaxies mentioned in Section~\ref{sec:nebred} for which the
Paschen-derived reddening is negligible.  Aside from these outliers,
the covering-fraction model successfully reproduces the reddening
computed from either the Balmer or Paschen lines for the vast majority
of galaxies in the sample.  


\subsubsection{Accounting for Diffuse Dust}
\label{sec:diffuse}

In the commonly-adopted ``two-component'' model for the dust
distribution, the nebular regions are subject to additional reddening
(and attenuation) relative to the stellar continuum, which is only
affect by a diffuse dust component \citep{charlot00}.  Along these
lines, in this section we briefly discuss a slight modification of the
covering-fraction model where both the reddened and unreddened
recombination line emission emerging from the vicinity of the $\hii$
regions is reddened by a diffuse distribution of dust with unity
covering fraction.  This type of configuration can be used to assess
the systematic uncertainty in $\ebmvneb^{\rm cov}$ and $\fcov$ that
may potentially arise from dust in the diffuse ISM.

Equation~\ref{eq:eq6} can be modified to include a second term that
accounts for a diffuse dust component described by the reddening of
the stellar continuum, $\ebmvcont$, and an associated dust attenuation
curve, $k_{\rm cont}(\lambda)$:
\begin{eqnarray}
R & = & \log_{10}\left[\frac{1-\fcov+\fcov\times 10^{-0.4\ebmvneb^{\rm cov}k_{\rm neb}(\lambda_1)}}{1-\fcov+\fcov\times 10^{-0.4\ebmvneb^{\rm cov}k_{\rm neb}(\lambda_2)}}\right]  \nonumber \\
&  & + \log_{10}\left[\frac{10^{-0.4\ebmvcont k_{\rm cont}(\lambda_1)}}{10^{-0.4\ebmvcont k_{\rm cont}(\lambda_2)}}\right].
\label{eq:diffuse}
\end{eqnarray}

\begin{figure}
  \epsscale{1.2}
  \includegraphics[width=1.0\linewidth]{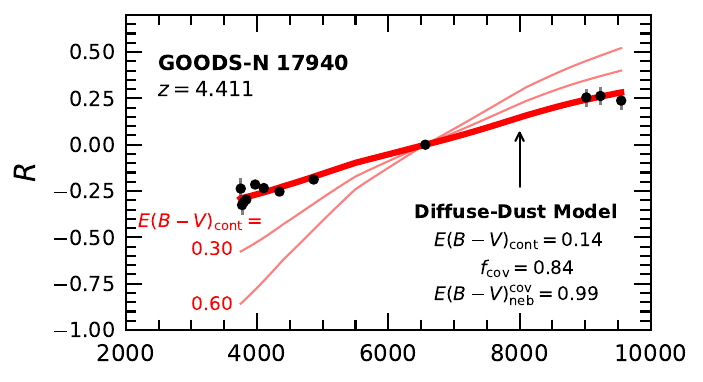}
    \caption{$R$ versus rest-frame wavelength for GOODSN-17940.  The
      measured $R$ are indicated by the black points.  The light red
      lines indicate the effect of diffuse dust reddening on $R$ with
      $\ebmvcont=0.30$ and $0.60$, assuming the fiducial
      covering-fraction model parameters of $\fcov = 0.91$ and
      $\ebmvneb^{\rm cov} = 0.84$ for this galaxy.  To match the data,
      a lower value of $\fcov = 0.84$ and a higher value of
      $\ebmvneb^{\rm cov}=0.99$ are required, assuming a diffuse dust
      reddening of $\ebmvcont = 0.14$ (thick red line).}
    \label{fig:diffusedemo}
\end{figure}

Figure~\ref{fig:diffusedemo} illustrates the effect of diffuse dust on
the wavelength dependence of $R$, in this case for GOODSN-17940.  For
this object, the fiducial covering-fraction modeling---which assumes
no diffuse dust---returns $\fcov = 0.91$ and $\ebmvneb^{\rm cov} =
0.84$.  Increasing the reddening of diffuse dust for the
aforementioned values of $\fcov$ and $\ebmvneb^{\rm cov}$ results in a
steeper relationship between $R$ and $\lambda_0$ (thin red lines).
Thus, matching the shallower observed dependence of $R$ on $\lambda_0$
with the diffuse-dust model requires that a larger fraction of the
recombination line fluxes are unreddened prior to encountering diffuse
dust.  This effect can be accomplished by either decreasing $\fcov$
and/or increasing $\ebmvneb^{\rm cov}$, as discussed in
Section~\ref{sec:subunityformalism}.  Indeed, if we assume a
diffuse-dust reddening of $\ebmvcont = 0.14$, as obtained from SED
fitting of GOODSN-17940 (Section~\ref{sec:sedfitting}), we find $\fcov
= 0.84\pm 0.03$ and $\ebmvneb^{\rm cov} = 0.99\pm 0.14$ (shown by the
thick red line in Figure~\ref{fig:diffusedemo}), values that are lower
and higher, respectively, compared to the values obtained when
assuming no diffuse dust.  Modeling all the galaxies in the sample
with a diffuse dust component results in $\fcov$ and $\ebmvneb^{\rm
  cov}$ that are on average $\approx 0.08$ lower and $\approx 0.17$
higher, respectively, compared to the values obtained with the
fiducial modeling which assumes no diffuse dust.

This modification to the fiducial covering-fraction model is useful
for quantifying the systematic uncertainty in $\fcov$ and
$\ebmvneb^{\rm cov}$ when a diffuse dust component is present.
However, it is likely that the same processes that yield a non-unity
covering fraction of dust proximate to the $\hii$ regions also results
in non-unity covering fraction of diffuse dust.  In this case, the
fiducial modeling may still provide useful approximations to the
covering fraction and reddening of dust affecting the recombination
lines.

\subsection{Normalization Constraints from the Covering-Fraction Modeling and the Total Nebular Attenuation Curve}
\label{sec:avecurvenorm}

An alternative method of constraining the normalization of the
effective attenuation curve is to require that intrinsic line fluxes
derived using this curve match those predicted by the
covering-fraction model (Section~\ref{sec:subunity}).  Using
Equation~\ref{eq:subunityeq1}, intrinsic $\hi$ emission line fluxes
were calculated for each galaxy based on the $\ebmvneb^{\rm cov}$ and
$\fcov$ presented in Section~\ref{sec:subunity}.  These intrinsic line
fluxes, along with the observed line fluxes, were used to calculate
the normalization of $k_{\rm neb}'(\lambda)$.  The ``effective'' $\rv$
computed in this manner vary in the range $\rv^{\rm eff} \simeq 3.2 -
16.4$, and typically lie between the $\rv$ computed using the
quadratic and linear long-wavelength extrapolations of the attenuation
curve.  

\begin{figure}
  \epsscale{1.15}
  \includegraphics[width=1.0\linewidth]{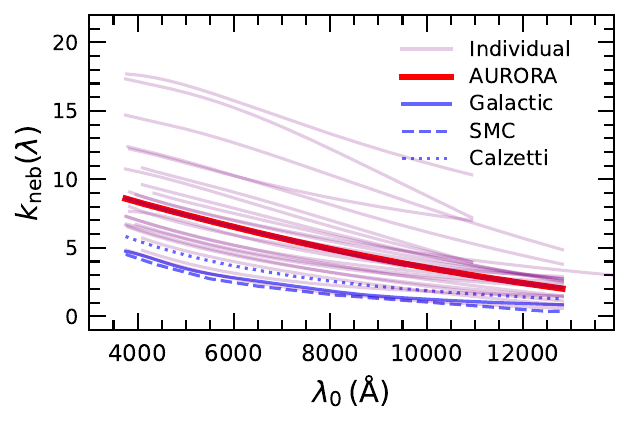}
    \caption{The average nebular attenuation curve, $k_{\rm
        neb}(\lambda)$, versus $\lambda_0$ for the main sample of 24
      galaxies (thick red curve).  The individual nebular dust
      attenuation curves are indicated by the light purple lines.  The
      curves for the Galactic, SMC, and Calzetti curves are shown by
      the solid, dashed, and dotted blue lines, respectively.}
   \label{fig:ktotal}
\end{figure}

Using the constraints on the intrinsic line fluxes from the
covering-fraction model, we find a median $\rv^{\rm eff} = 6.957$ for
the 24 galaxies.  Combining this with the relative attenuation curve
(Equation~\ref{eq:effcurveshape}), we obtain the following for the
total (average) nebular attenuation curve:
\begin{eqnarray}
\langle k_{\rm neb}(\lambda)\rangle & = & -7.241 + \frac{17.002}{\lambda/\mu{\rm m}} - \frac{8.086}{(\lambda/\mu{\rm m})^2} \nonumber \\
& & + \frac{2.177}{(\lambda/\mu{\rm m})^3} - \frac{0.319}{(\lambda/\mu{\rm m})^4} + \frac{0.021}{(\lambda/\mu{\rm m})^5},
\label{eq:avecurve}
\end{eqnarray}
again valid over the wavelength range $0.35\la \lambda \la
1.28$\,$\mu$m.  The average and individual effective nebular
attenuation curves are compared with other common extinction and
attenuation curves in Figure~\ref{fig:ktotal}.  As noted before, the
average nebular attenuation curve has a shape similar to that of other
common extinction and attenuation curves at $\lambda\la 5500$\,\AA.
However, the large difference in normalization between the average
curve derived for the AURORA sample and the Galactic, SMC, and
Calzetti curves becomes apparent only when the Paschen line fluxes are
considered jointly with the Balmer line fluxes.  

\subsection{Comparison to the \citet{reddy20} Nebular Attenuation Curve}
\label{sec:comparer20}

\citet{reddy20} used ground-based optical spectroscopic data from the
MOSFIRE Deep Evolution Field (MOSDEF) survey \citep{kriek15} to
provide the first constraints on the shape of the nebular attenuation
curve at $z\sim 2$.  Based on the first five low-order Balmer lines
detected in the composite spectrum of 532 star-forming galaxies
accessible in that study, they found a nebular attenuation curve that
closely resembles other common extinction and attenuation curves,
including the Galactic extinction curve of \citet{cardelli89}, within
the uncertainties.  As noted in Section~\ref{sec:indandave}, the
average nebular attenuation curve found in this work also has a shape
similar to that of the Galactic extinction curve at $\lambda \la
5500$\,\AA, consistent with the shape found by \citet{reddy20}.  The
key difference, however, is that this analysis indicates a
substantially higher $\rv=6.96$ than that of the Galactic extinction
curve ($\rv \simeq 3.1$), a result that only becomes apparent with the
inclusion of longer-wavelength, less-reddening-sensitive near-IR
Paschen line measurements made possible by JWST, and which were
unavailable for the MOSDEF sample.  For most galaxies in the sample,
the Paschen-only line ratios, as well as those formed from a
combination of Paschen and Balmer lines, suggest the presence of star
formation that is heavily reddened in the Balmer lines, leading to a
higher $\rv$ than the Galactic extinction curve.

\subsection{Physical Interpretation of the Normalization of the Effective Attenuation Curve}
\label{sec:rvdiscuss}

\begin{figure}
  \epsscale{1.15}
  \includegraphics[width=1.0\linewidth]{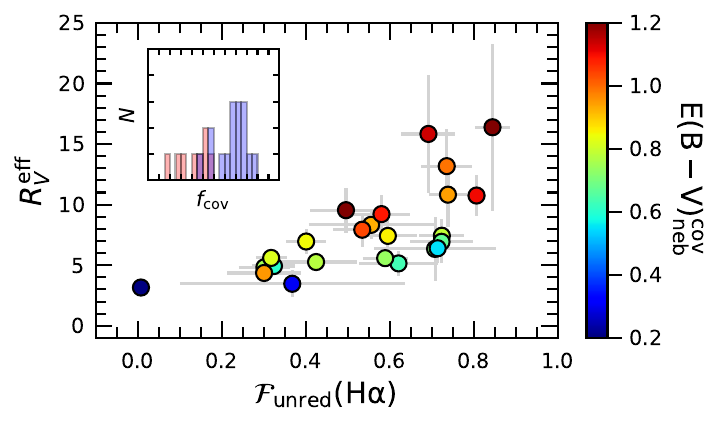}
    \caption{$\rv^{\rm eff}$ as a function of the unreddened fraction
      of light contributing to the observed $\ha$ line flux, $\mathcal{F}_{\rm
        unred}(\ha)$, for the 24 galaxies in the sample.  The points
      are color-coded by $\ebmvneb^{\rm cov}$ as determined from the
      covering-fraction model.  The inset panel shows the histogram of
      $\fcov$ for galaxies with $\mathcal{F}_{\rm unred}(\ha)>0.7$ (red) and
      $\mathcal{F}_{\rm unred}(\ha)<0.7$ (blue).}
   \label{fig:fobsrv}
\end{figure}

In the covering-fraction model (Section~\ref{sec:subunity}), the
geometry of dust and stars is encapsulated by the covering fraction of
dust ($\fcov$) and reddening ($\ebmvneb^{\rm cov}$), which
parameterize the variation in optical depths along the sightlines to
OB associations.  The unreddened fraction of the observed line flux at
a given wavelength predicted by this model (see
Section~\ref{sec:subunity} and Figure~\ref{fig:unredfrac}) should
correlate with $\rv$, allowing us to connect the normalization of the
effective attenuation curve to the physically motivated
covering-fraction model.  As shown in Figure~\ref{fig:fobsrv} for the
24 galaxies in the present sample, $\rv$ progressively increases as
the fraction of unreddened light contributing to the observed $\ha$
flux rises.  From a Spearman test, the probability of a null
correlation between $\rv$ and $\mathcal{F}_{\rm unred}(\ha)$ is
$p=5.6\times 10^{-5}$, and remains highly significant ($p<0.0008$)
regardless of the method used to constrain $\rv$.  This holds true
whether $\rv$ is determined using a quadratic or linear extrapolation,
or by forcing the intrinsic line fluxes to match those predicted by
the covering-fraction model, as all approaches yield $\rv$ that
correlate strongly with each other.

Recall that $\mathcal{F}_{\rm unred}(\lambda)$ may be elevated either
due to a low $\fcov$ or a high $\ebmvneb^{\rm cov}$ (see
Section~\ref{sec:subunityformalism}).  Our analysis suggests that both
factors are at play.  The inset panel in Figure~\ref{fig:fobsrv}
presents histograms of $\fcov$ for galaxies above and below
$\mathcal{F}_{\rm unred}(\ha) = 0.7$.  Galaxies with $\mathcal{F}_{\rm
  unred}(\ha) > 0.7$ show a systematically lower $\fcov$ (with a
median $\fcov = 0.72$) compared to galaxies with $\mathcal{F}_{\rm
  unred}(\ha) < 0.7$ (with a median $\fcov = 0.90$).  Moreover, the
color-coding of points in Figure~\ref{fig:fobsrv} demonstrates that
galaxies exhibiting higher $\mathcal{F}_{\rm unred}(\ha)$ also tend to
have systematically redder $\ebmvneb^{\rm cov}$.

\begin{figure}
  \epsscale{1.15}
  \includegraphics[width=1.0\linewidth]{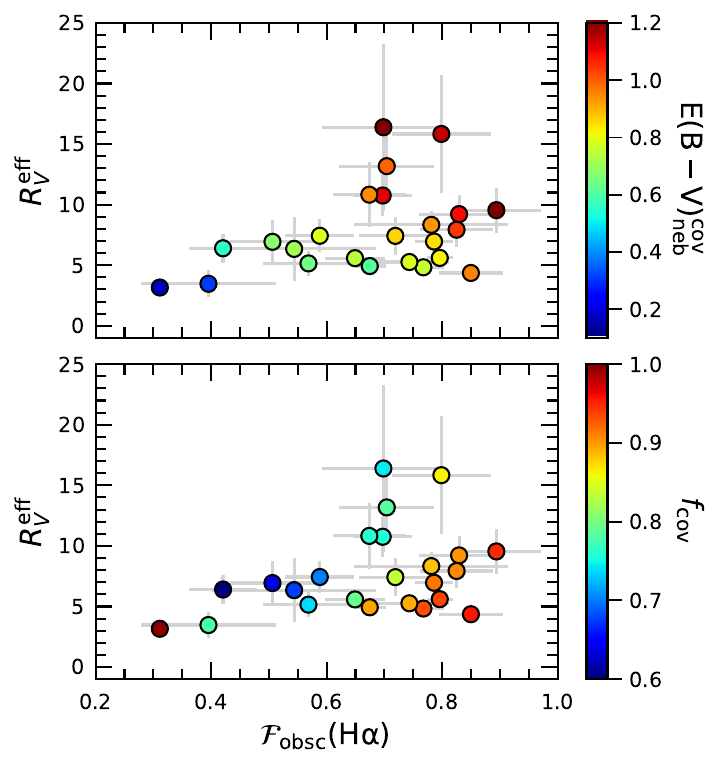}
    \caption{$\rv^{\rm eff}$ as a function of the fraction of
      dust-obscured $\ha$ luminosity, $\mathcal{F}_{\rm obsc}(\lambda)$, as
      given in Equation~\ref{eq:obscuredfrac}.  The symbols in the top
      and bottom panels are color coded by $\ebmvneb^{\rm cov}$ and
      $\fcov$, respectively, as determined from the covering-fraction
      model.}
   \label{fig:fobscuredrv}
\end{figure}

The relationship between $\rv$ and $\mathcal{F}_{\rm unred}(\lambda)$, along
with the scatter in this relationship, provides valuable insights into
the underlying factors that contribute to variations in $\rv$ within
the sample.  In particular, the correlation of this scatter with
$\ebmvneb^{\rm cov}$ and $\fcov$ implies that these two factors are
key drivers of the observed variation in $\rv$.  Given the considerable
interest in understanding how attenuation curves vary in the dustiest
galaxies, we can further investigate how $\rv$ changes with respect to
the fraction of intrinsic luminosity that is obscured, i.e., $\mathcal{F}_{\rm
  obsc}(\lambda)$ (Equation~\ref{eq:obscuredfrac}), as shown in
Figure~\ref{fig:fobscuredrv}.  Galaxies with the highest values of
$\rv$ in the sample (e.g., $\rv \ga 10$) tend to be more ``dusty,''
where a larger fraction of the intrinsic $\ha$ luminosity is obscured
by dust.  Conversely, the dustiest galaxies in the sample (e.g., those
with $\mathcal{F}_{\rm obsc}(\ha) \ga 0.70$) exhibit a range of $\rv$ values
that depend on the dust-covering fraction, $\fcov$, and $\ebmvneb^{\rm
  cov}$.  Specifically, dusty galaxies with modest dust-covering
fractions and high $\ebmvneb^{\rm cov}$ tend to have higher $\rv$,
while those with higher $\fcov$ and modest values of $\ebmvneb^{\rm
  cov}$ tend to have lower $\rv$.  These results suggest that dusty
galaxies may exhibit a range of $\rv$ depending on the actual covering
fraction of---and reddening provided by---the dust.

To conclude, the normalization of the nebular attenuation curve shows
a positive correlation with the fraction of unobscured light
contributing to the observed line fluxes.  Galaxies with high $\rv$
values systematically exhibit lower values of $\fcov$ and higher
$\ebmvneb^{\rm cov}$.  Although this analysis specifically addresses
the nebular attenuation curve, the general idea that a greater
fraction of unobscured sightlines (low $\fcov$) mixed with
heavily-reddened (high $\ebmv$) sightlines leads to a higher $\rv$
supports the general paradigm outlined for the reddening curves
of the {\em stellar continuum} in young star-forming galaxies, where
an increase in unobscured sightlines typically results in flatter (or
higher $\rv$) attenuation curves (e.g., \citealt{witt96, calzetti00,
  narayanan18, trayford20}).  Thus, examining the variation in $\rv$
for nebular attenuation curves can provide insights into the factors
that may contribute to variations in $\rv$ for stellar attenuation
curves.

Since the AURORA sample is broadly representative of galaxies at
$z\sim 1.5-4.5$, the average nebular attenuation curve given by
Equation~\ref{eq:avecurve} should be applicable to statistical samples
of star-forming galaxies at these redshifts, and potentially even
higher redshifts as well (Section~\ref{sec:highzcurve}).  However,
caution is advised when applying this curve to galaxies whose
properties differ significantly from those in the AURORA sample, such
as those with $\mathcal{F}_{\rm unred}$ values near the extremes of
the distribution shown in Figure~\ref{fig:fobsrv}.  In such
cases---i.e., where the covering fraction is high, or where the
covering fraction is low and the reddening is high---it may be more
appropriate to assume a lower or higher $\rv$ than the one associated
with the curve given in Equation~\ref{eq:avecurve}.  The key
conclusion of this analysis is that the diversity of shapes and
normalizations of the effective nebular attenuation curves presented
in Section~\ref{sec:effcurve} can be accommodated by assuming the
Galactic extinction curve---and therefore similar dust grain
properties as seen along Milky Way sightlines---with a subunity
covering fraction of that dust (Section~\ref{sec:subunity}).

\section{Conclusions}
\label{sec:conclusions}

We used deep JWST/NIRSpec $\lambda = 1 - 5$\,$\mu$m spectroscopy from
the AURORA survey to measure numerous $\hi$ Balmer and Paschen
recombination emission lines in 24 individual star-forming galaxies at
$z=1.52-4.41$.  The measurements were used to derive the nebular
attenuation curves and nebular reddening in these galaxies.  To
interpret the multiple recombination line ratios, we also considered
a model that departs from the usual assumption of a uniform screen of
dust.  The primary results from our analysis are as follows:

\begin{itemize}

\item For at least half the galaxies in the sample, the nebular
  reddening derived from the Paschen lines exceeds the value obtained
  from the Balmer lines, adopting the usual assumptions of a uniform
  screen ($100\%$ covering fraction) of dust and the Galactic
  extinction curve (Section~\ref{sec:nebred} and
  Figure~\ref{fig:reddening}).  Other standard extinction and
  attenuation curves in the literature yield similar results.  We
  interpret the systematic offset between the Paschen- and
  Balmer-inferred reddening as an indication of the presence of star
  formation that is heavily reddened in the Balmer lines.

\item We used the $\hi$ recombination line ratios to calculate the
  effective nebular attenuation curves of the individual galaxies
  (Section~\ref{sec:effmeth}).  The derived attenuation curves are
  similar in shape to those of standard extinction and attenuation
  curves at $\lambda \la 6000$\,\AA\, (Section~\ref{sec:indandave}).
  However, at longer wavelengths, the individual nebular attenuation
  curves for most galaxies diverge from standard extinction and
  attenuation curves (Figure~\ref{fig:kquadprime_examples}), a
  behavior that is also reflected in the average nebular dust
  attenuation curve of the 24 galaxies
  (Figure~\ref{fig:kquadprime_average}).  We do not find substantial
  evidence for any redshift evolution in the shape of the average
  nebular dust attenuation curve based on the limited data available
  for $z>3$ galaxies in the AURORA sample (Section~\ref{sec:highzcurve} and
  Figure~\ref{fig:kquadprime_highz}).

\item The $\rv$ (or normalizations) of the effective nebular
  attenuation curves are in the range $\rv\simeq 3.2-16.4$,
  considerably higher on average than the values found in a range of
  environments in the Milky Way (Figure~\ref{fig:rvcomphist}),
  suggesting that the geometry of dust with respect to the stars
  (i.e., the dust covering fraction, $\fcov$) is driving the variation
  in $\rv$ (Sections~\ref{sec:rvnorm} and \ref{sec:avecurvenorm}).

\item Based on the offsets between the Balmer and Paschen-inferred
  reddening and the large $\rv$ values, both of which suggest the
  importance of the dust-stars geometry on the effective nebular
  attenuation curves, we explored a model that parameterizes
  variations in the dust optical depths towards OB associations by
  assuming a sub-unity covering fraction of dust
  (Section~\ref{sec:subunity}).  By fitting this model to the observed
  $\hi$ recombination line ratios, we obtained dust covering fractions
  in the range $\fcov \simeq 0.6 - 1.0$ and line-of-sight reddening
  values of $\ebmvneb^{\rm cov} \simeq 0.2 - 1.2$
  (Figure~\ref{fig:subunity_parms}).  Thus, our analysis suggests that
  the diversity of effective nebular attenuation curves can be
  accounted for by assuming dust grain properties similar to that of
  Galactic sightlines (i.e., the Galactic extinction curve), but with
  a subunity covering fraction of dust.  The light-weighted reddening
  at the wavelengths of the Balmer and Paschen lines computed using
  the covering-fraction model agree well with the corresponding values
  based purely on the Balmer lines or Paschen lines
  (Figure~\ref{fig:ebmvlw}).

\item Further, we find a strong correlation between $\rv$ and the
  fraction of unreddened light contributing to the observed $\hi$ line
  fluxes (Section~\ref{sec:rvdiscuss} and Figure~\ref{fig:fobsrv}).
  Thus, galaxies with unreddened sightlines mixed with heavily-reddened
  sightlines have higher $\rv$ values.  Dusty galaxies may exhibit a
  range of $\rv$ depending on the covering fraction of, and reddening
  provided by, the dust (Figure~\ref{fig:fobscuredrv}).

\end{itemize}

The AURORA sample includes typical star-forming galaxies at $z\sim
1-4$, with bolometric luminosities and dust obscurations lower than
those of the rarer ultraluminous infrared (and brighter) galaxies that
are routinely detected individually with ALMA and other far-IR
facilities.  Yet, even within these typical galaxies of the AURORA
sample, we find evidence of heavily-reddened star formation.  Though
the distribution of dust towards OB associations likely varies on
parsec scales, which are too small to resolve for unlensed galaxies
even with JWST, our analysis demonstrates that even integrated $\hi$
recombination line ratios can provide valuable constraints on the
covering fraction of dust and dust-obscured luminosity in these
galaxies.  

This analysis points to a number of other investigations to further
exploit and leverage the exquisite JWST spectroscopy of high-redshift
galaxies.  While our analysis has focused on the nebular dust
attenuation curves of individual galaxies, there are many unexplored
avenues that can be pursued by constructing composite spectra in bins
of galaxy properties, particularly those which are known to correlate
with dust attenuation (e.g., UV slope, stellar mass, gas-phase
metallicity, etc.).  This will allow us to better understand the
factors that modulate the shape and normalization of the nebular dust
attenuation curve, including possible variations in the dust grain
composition and size distribution.  Earlier studies to constrain the
shape of the dust attenuation curve of the {\em stellar continuum}
(e.g., \citealt{reddy15, shivaei20b}) can be improved upon with the
precise measurements of nebular reddening enabled with JWST/NIRSpec,
combined with more extensive and deeper near-IR photometry available
with JWST/NIRCam.

Well-constructed composite spectra may yield additional insights into
the physical conditions (e.g., level of $l$-mixing, continuum pumping)
modulating the $\hi$ recombination line fluxes, in particular by
focusing on the higher-order Balmer and Paschen line ratios that are
close enough in wavelength that their observed ratios should be
similar to the intrinsic ones (e.g., Appendix~\ref{sec:intrinsic}).
Separately, more precise constraints on $\rv$ can be attained with a
similar analysis for galaxies at lower redshifts ($z<1.67$;
Figure~\ref{fig:redshifts}) where Pa$\alpha$ (Pa4) is accessible
within the JWST/NIRSpec wavelength range.


The covering-fraction model assumes a simple two-component
configuration in which the observed recombination line flux is the sum
of an unreddened component and a component reddened by a single value.
This approach is analogous to the formalism used to derive gas
covering fractions from fitting rest-frame UV spectra (e.g.,
\citealt{reddy16a, reddy16b, steidel18, gazagnes20}).  While this
model provides a good approximation for galaxies in our
sample---successfully reproducing all the $\hi$ line ratios with only
one additional free parameter ($f_{\rm cov}$)---the true reddening
distribution within galaxies is likely more complex.  Integrated line
measurements alone do not provide enough power to constrain more
complicated models that include multiple reddened components or a
continuous distribution of reddening.  Spatially resolved
line-emission observations (e.g., with IFU spectroscopy) can be used
to explore how such reddening distributions map onto the two
parameters constrained by the covering-fraction model ($\ebmvneb^{\rm
  cov}$ and $f_{\rm cov}$).

The novel constraints on dust covering fractions towards OB
associations may be particularly relevant for understanding several
key aspects of high-redshift galaxies.  The dust covering fraction
likely correlates with gas covering fraction, which in turn has been
shown to be a key factor in the escape of ionizing radiation, a topic
of importance for studies of reionization (e.g.,
\citealt{zackrisson13, rivera15, trainor15, reddy16b, steidel18,
  gazagnes18, jaskot19, reddy22}).  Thus, it is important to investigate the
connection between dust covering fraction deduced from the $\hi$
recombination lines and proxies for $\hi$ gas and metal covering
fractions---e.g., from the depths of the Lyman series absorption lines
\citep{reddy16b, steidel18, gazagnes18} or the equivalent widths of
saturated low-ionization interstellar absorption lines
\citep{shapley03, berry12, jones12, reddy16b, du18, pahl20, reddy22}.
Connecting these covering fractions to properties that may indicate
the efficiency of stellar feedback (e.g., wind velocities, SFR surface
densities, etc.) will yield valuable insights into the effect of such
feedback on the porosity of the ISM and the escape of Ly$\alpha$ and
Lyman continuum photons.

The broader implications of the nebular dust attenuation curves
presented here, including their impact on dust-corrected line
luminosities and line ratios, nebular-based SFRs, and differential
reddening of the nebular lines and stellar continuum, are presented in
Paper II (Reddy et~al., submitted).  Additionally, Paper II discusses
the calibration between JWST/MIRI dust emission measurements and total
SFRs in the context of the $\hi$ recombination line analysis.  It also
presents a joint analysis of the nebular and stellar reddening curves
for one of the youngest galaxies in the AURORA sample, GOODSN-17940.

\begin{acknowledgements}

This work is based on observations made with the NASA/ ESA/CSA James
Webb Space Telescope. The data were obtained from the Mikulski Archive
for Space Telescopes at the Space Telescope Science Institute, which
is operated by the Association of Universities for Research in
Astronomy, Inc., under NASA contract NAS5-03127 for JWST. The specific
observations analyzed can be accessed via doi:10.17909/hvne7139. We
also acknowledge support from NASA grant No. JWST-GO-01914.  Some of
the data products used in this analysis were retrieved from the Dawn
JWST Archive (DJA). DJA is an initiative of the Cosmic Dawn Center
(DAWN), which is funded by the Danish National Research Foundation
under grant DNRF140.

\end{acknowledgements}

\facility{{\em JWST}/NIRSpec}



\appendix

\section{Variations in the Intrinsic Line Ratios with Electron Temperature and Assumed Level of $l$-Mixing}
\label{sec:intrinsic}

The default values of the intrinsic line ratios adopted in this
analysis were obtained from PyNeb, assuming $n_{\rm e} =
100$\,cm$^{-3}$ and $T = 15,000$\,K (Table~\ref{tab:intrinsic}).  In
the following, we explore how different assumptions of the $\hii$
region temperature and level of $l$-mixing affect these line ratios,
and the impact on several of the most important quantities discussed
in this analysis.  To aid the following discussion, it is useful to
define the parameter, $f_{\rm lr}$, as the fractional variation in an
intrinsic line ratio relative to the default value listed in
Table~\ref{tab:intrinsic}:
\begin{eqnarray}
f_{\rm lr} & = & \left(\frac{f_0(\lambda)}{f_0(\ha)}\right) \Big/ \left(\frac{f_0(\lambda)}{f_o(\ha)}\right)_{\rm def},
\end{eqnarray}
where the denominator is the default intrinsic line ratio.

\begin{figure}
  \epsscale{1.0}
  \includegraphics[width=1.0\linewidth]{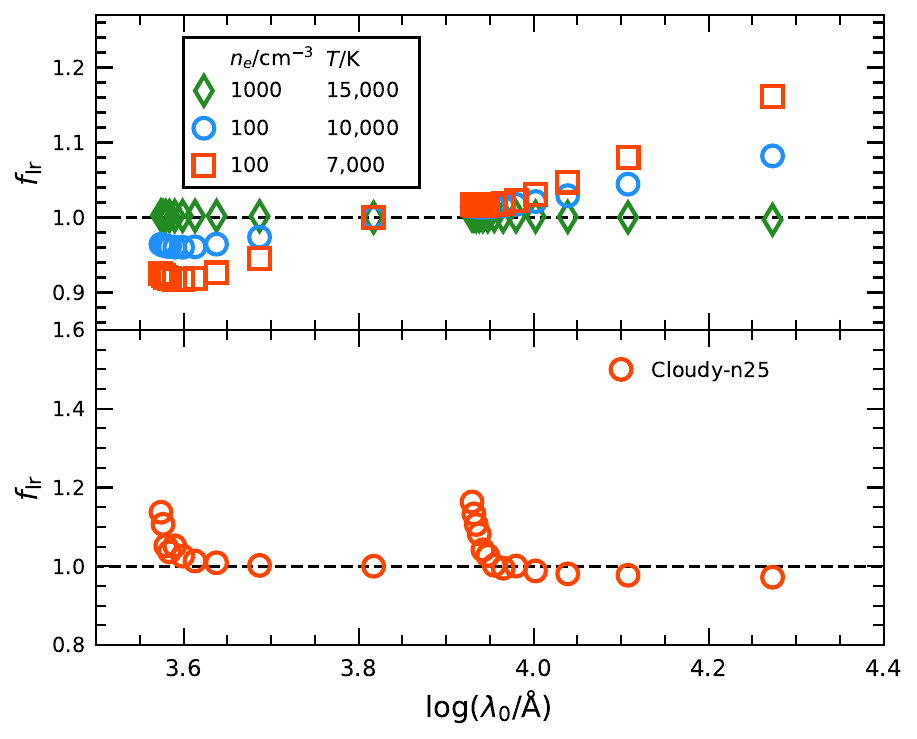}
    \caption{Fractional variation in the intrinsic line ratios
      relative to the default values reported by PyNeb and listed in
      Table~\ref{tab:intrinsic}, which assume $n_e=100$\,cm$^{-3}$ and
      $T=15,000$\,K.  The top panel shows the effect of increasing the
      electron density by an order of magnitude ($n_e =
      1000$\,cm$^{-3}$ at the same temperature as the default model
      (green diamonds).  Also shown are two other models with $n_e =
      100$\,cm$^{-3}$ and lower temperatures of $T=10,000$\,K and
      $7,000$\,K (blue circles and red squares, respectively).  The
      bottom panel shows the fractional variation in the intrinsic
      line ratios for different assumptions of $l$-mixing, computed
      using Cloudy modeling (see text). }
    \label{fig:intrinsic}
\end{figure}

The top panel of Figure~\ref{fig:intrinsic} shows $f_{\rm lr}$ for
three combinations of $n_e$ and $T$, based on the values reported by
PyNeb.  For the range of $n_e$ inferred for high-redshift star-forming
galaxies ($n_e \la 1000$\,cm$^{-3}$), there is little variation in the
intrinsic line ratios at a fixed $T$.  For the Paschen lines, density
effects become important at $n_e \ga 10^5$\,cm$^{-3}$, which causes
these lines to be weaker relative to the Balmer lines
\citep{ferguson97}.  For the Balmer lines, density effects only become
important closer to the critical density of $\simeq 10^8$\,cm$^{-3}$.
In contrast, variations in the electron temperature within the range
found for $\hii$ regions can cause non-negligible shifts in the
higher-order Balmer and Paschen lines.  For example, for
high-metallicity galaxies with $T\approx 7,000$\,K, the H12-to-$\ha$
ratio is $\approx 0.92\%$ of the default value.  Qualitatively, higher
temperatures (corresponding to lower metallicity $\hii$ regions) lead
to a larger fraction of electrons occupying higher-$n$ states,
resulting in an increased emissivity of the high-order lines
relative to the low-order lines of the same series.  


\begin{figure}
  \epsscale{1.0}
  \includegraphics[width=1.0\linewidth]{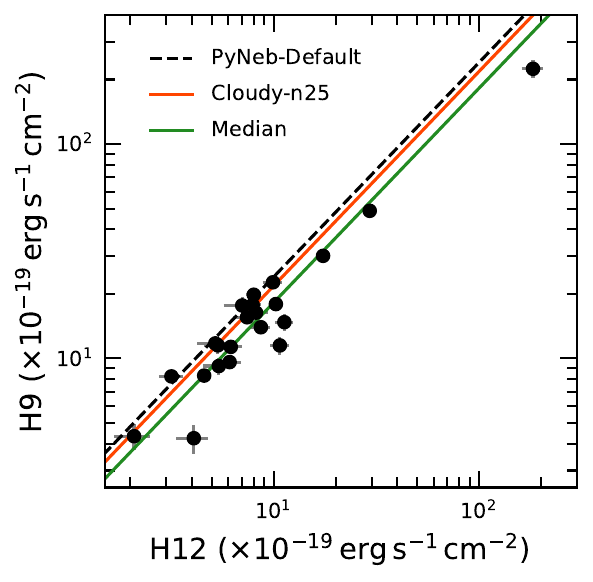}
    \caption{Comparison of the H9 and H12 fluxes of 22 galaxies in the
      parent AURORA sample with $>5\sigma$ detections of both lines.
      The dashed line indicates the expected intrinsic ratio of H9/H12
      as reported by PyNeb using the default values of $n_e =
      100$\,cm$^{-3}$ and $T=15,000$\,K.  The intrinsic ratios
      calculated with the Cloudy-n25 model described in the text is
      shown by the red line.  The green line denotes the observed
      median ratio of H9/H12.}
    \label{fig:h9h12}
\end{figure}

\begin{figure*}
  \epsscale{1.0}
  \includegraphics[width=1.0\linewidth]{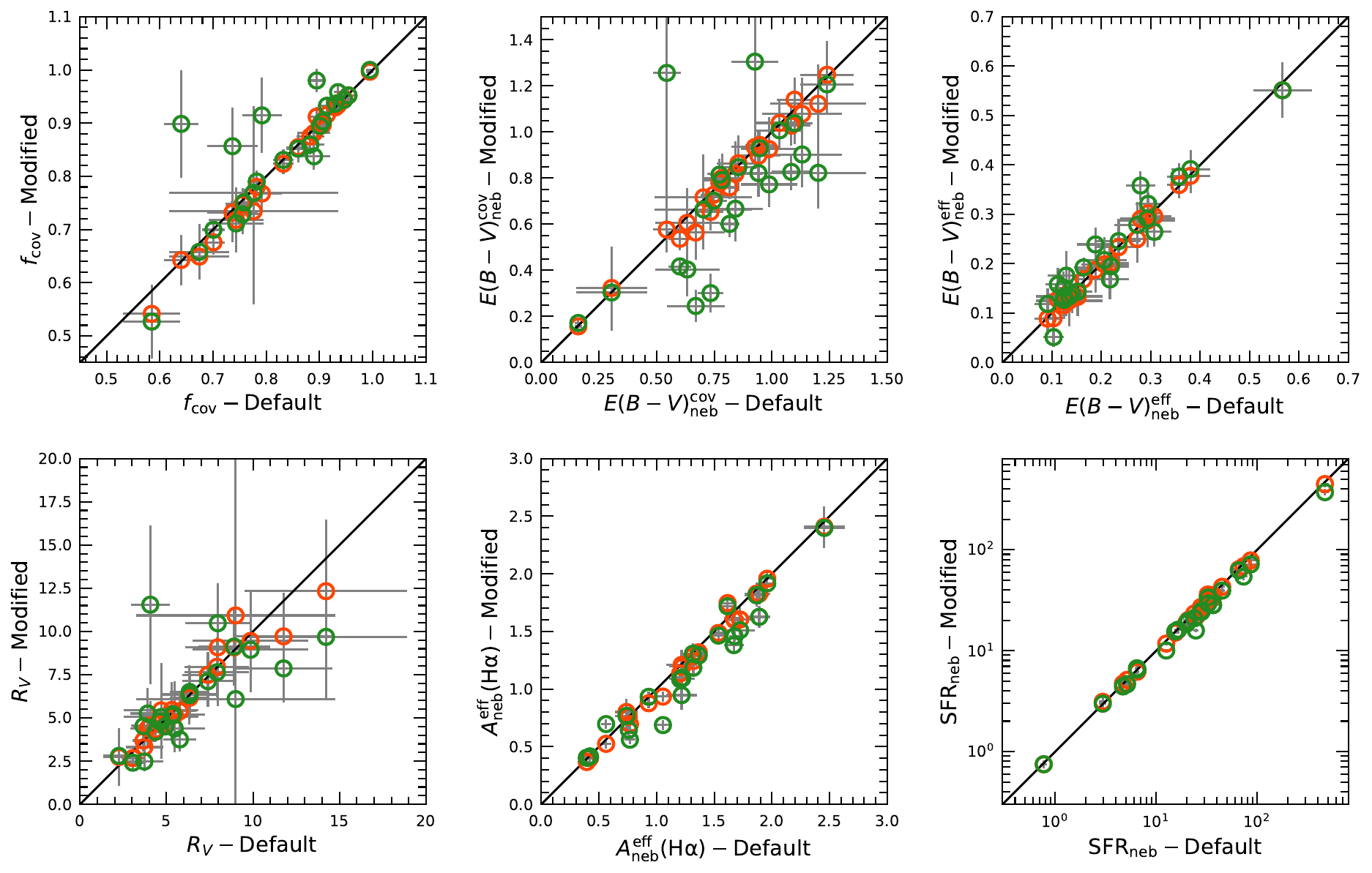}
    \caption{Effect of adjusting the intrinsic $\hi$ recombination
      line ratios on several quantities analyzed in this work,
      including $\fcov$, $\ebmvneb^{\rm cov}$, $\ebmvneb^{\rm eff}$,
      $\rv$, $A_{\rm neb}^{\rm eff}(\ha)$, and $\sfrneb$.  In each
      panel, the open red and blue circles indicate values obtained
      with the intrinsic ratios from the Cloudy-default and PyNeb
      $T=10,000$\,K models (see text).}
    \label{fig:intrinsiceffect}
\end{figure*}

The $\hi$ line emissivities reported by PyNeb are taken from
\citet{hummer87} and \citet{storey95}, who present the most definitive
calculations of these emissivities for a 1000-level H atom,
considering all angular momentum ($l$) states.  \citet{pengelly64}
calculate, as a function of density, the $n$ level above which the
collisional rate exceeds the radiative lifetime, such that
perturbations from collisions cause electrons to mix between different
angular momentum states, a process known as $l$-mixing.  This in turn
alters the transition probabilities and resulting line emissivities.
For instance, these authors calculate that all levels with $n\ga 25$ are
well mixed for $n_e \sim 100$\,cm$^{-3}$.  

The bottom panel of Figure~\ref{fig:intrinsic} shows $f_{\rm lr}$ for
intrinsic line ratios computed from Cloudy v23.01 by invoking an H
atom with 25 $l$-resolved levels and an additional 300 ``collapsed''
levels that are well mixed, or thermalized (the ``Cloudy-n25'' model).
In the bottom panel, the default line ratios used to define $f_{\rm
  rl}$ are taken from PyNeb for $T = 10,000$\,K---which is close to
the average value computed by Cloudy over all ionization zones of the
$\hii$ region---and $n_e = 100$\,cm$^{-3}$.  As expected, the largest
discrepancies in the line ratios occur for the highest-order Balmer
and Paschen lines, as these high $n$-levels are more significantly
affected by the degree of $l$-mixing.

It is useful to examine whether the observed line ratios obtained with
AURORA are consistent with the predicted intrinsic line ratios.
Specifically, H12 and H9 are far enough apart in wavelength that their
ratio can provide some leverage on the models shown in
Figure~\ref{fig:intrinsic}, yet they are still close enough in
wavelength that their observed line ratio should be close to the
intrinsic value.  A comparable analysis can be applied to the
high-order Paschen lines (e.g., Pa14 and higher), although these lines
are relatively weaker and thus provide less constraining power.

Figure~\ref{fig:h9h12} displays the H9 and H12 fluxes of the 22
galaxies in the AURORA parent sample that have $>5\sigma$ detections
of both lines.  The intrinsic ratio of H9-to-H12 is indicated by the
dashed and solid lines for the default PyNeb values adopted in this
analysis and for the Cloudy model discussed above, respectively.
Interestingly, the observed H9/H12 ratios for the majority of galaxies
are systematically and significantly lower than the intrinsic values
reported by PyNeb and Cloudy-n25 (varying the temperature does little
to alter the intrinsic ratio), in the opposite direction one would
expect if the ratio is modified by dust reddening.  The observed median
ratio of H9/H12 is indicated by the green line.
A similar conclusion is reached if only the 6 objects with $>10\sigma$
detections of both lines are considered: these objects have the
highest equivalent widths, $W_\lambda$, of the Balmer lines where
systematic uncertainties in the stellar Balmer absorption corrections
are minimal, and where possible variations in the continuum fit under
the lines should not materially affect the line ratios.  The
offsets between the observed and intrinsic ratios may indicate Case A
conditions---which result in a lower intrinsic H9/H12 ratio than the
Case B prediction---or possibly different degrees of $l$-mixing or
continuum-pumping of the $\hi$ lines through stellar absorption (e.g.,
see \citealt{mesadelgado09, dominguezguzman22} for similar
observations in the Orion Nebula and $\hii$ regions in the Magellanic
Clouds, respectively).


Whatever processes may be at play in modulating the H12 and H9 line
emissivities it is useful to examine how altering the intrinsic ratios
affects several of the key quantities discussed in this analysis,
including $\fcov$, $\ebmvneb^{\rm cov}$, $\ebmvneb^{\rm eff}$, $\rv$,
$A_{\rm neb}^{\rm eff}$, and $\sfrneb$.
Figure~\ref{fig:intrinsiceffect} shows the comparison of the values
derived with the default PyNeb assumption ($T=15,000$\,K) and those
obtained with the Cloudy-n25 intrinsic ratios (red circles).
Additionally, the green circles denote the values obtained if we fix
the intrinsic ratios of H9/H10, H10/H11, and H11/H12 to the median
observed values.  Though there are a few outliers for some of the
parameters (e.g., $\fcov$ and $\ebmvneb^{\rm cov}$), for the vast
majority of galaxies in the sample, there are no large systematic
offsets larger than the measurement uncertainties. For the most part,
there is a one-to-one correspondence in the variables derived using
different sets of intrinsic line ratios, such that the ordering of
objects in parameter space is preserved (e.g., those with high $\fcov$
using the default intrinsic line ratios have similarly high $\fcov$
using the modified intrinsic line ratios).  Thus, all of the
correlations discussed in the analysis are preserved regardless of
which sets of intrinsic line ratios are assumed.

\section{Reproducibility of Line Ratios assuming Unity and Sub-unity Covering Fractions}
\label{sec:reproducibility}

\begin{figure*}
  \epsscale{1.2}
  \includegraphics[width=1.0\linewidth]{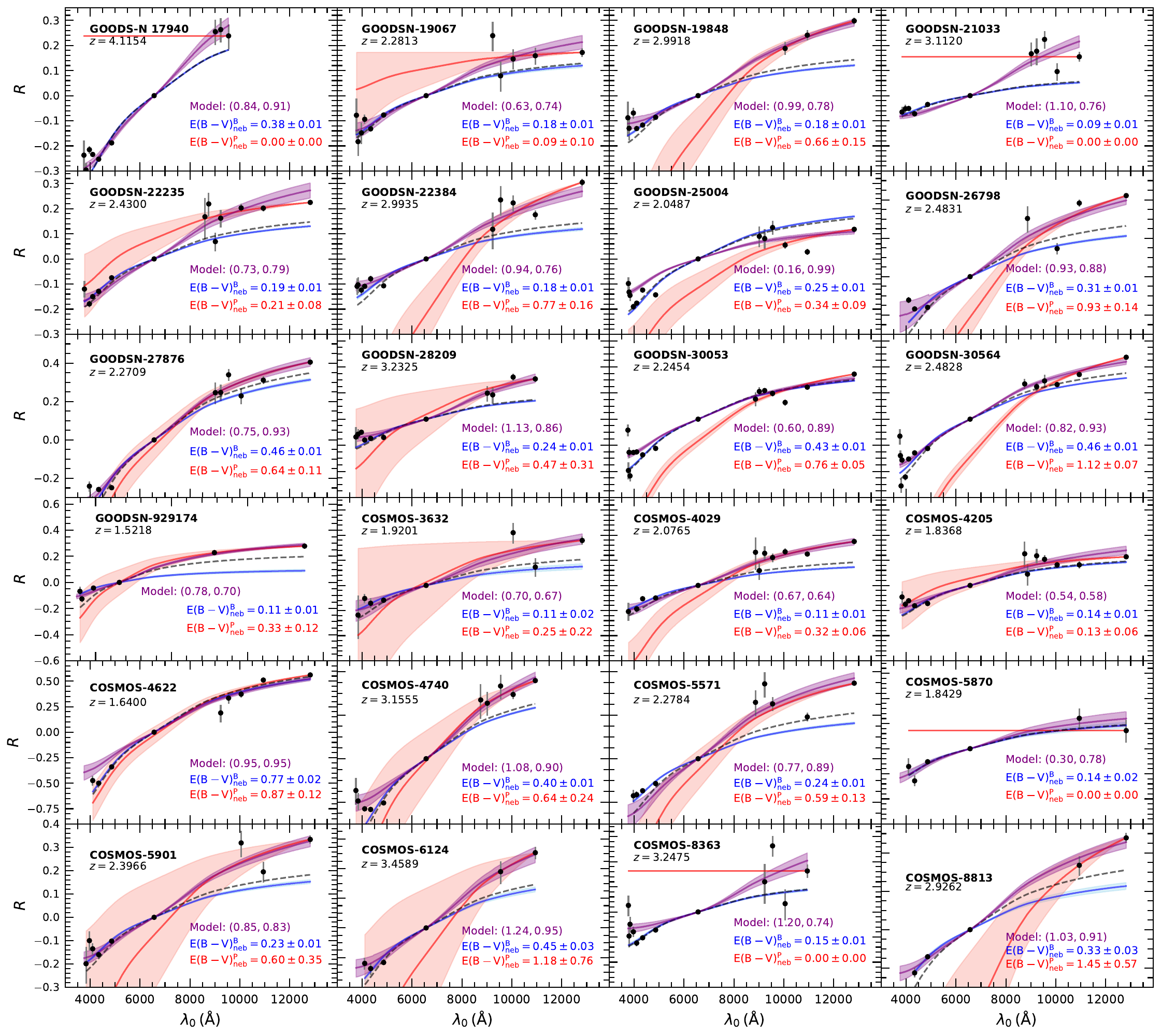}
     \caption{Same as Figures~\ref{fig:examples} and
       \ref{fig:subunity_examples} for the full sample of 24 galaxies.
       The best-fit covering fraction model parameters are indicated
       as a pair in each panel: $(\ebmvneb^{\rm cov},f_{\rm cov})$.}
     \label{fig:r_subplots}
\end{figure*}

Figure~\ref{fig:r_subplots} is the same as
  Figure~\ref{fig:examples} and \ref{fig:subunity_examples}.  The
  panels show fits to $R$ vs $\lambda_0$ for each of the 24 galaxies
  in the sample, assuming a unity covering fraction and the
  \citet{cardelli89} extinction curve---blue, red, and dashed lines
  for fits to the Balmer lines, Paschen lines, and all lines,
  respectively.  The purple curves indicate the fits assuming a
  sub-unity covering fraction of dust.

\section{Testing for Biases in the Covering-Fraction Model}
\label{sec:covfracbias}

\begin{figure}
  \epsscale{1.2}
  \includegraphics[width=1.0\linewidth]{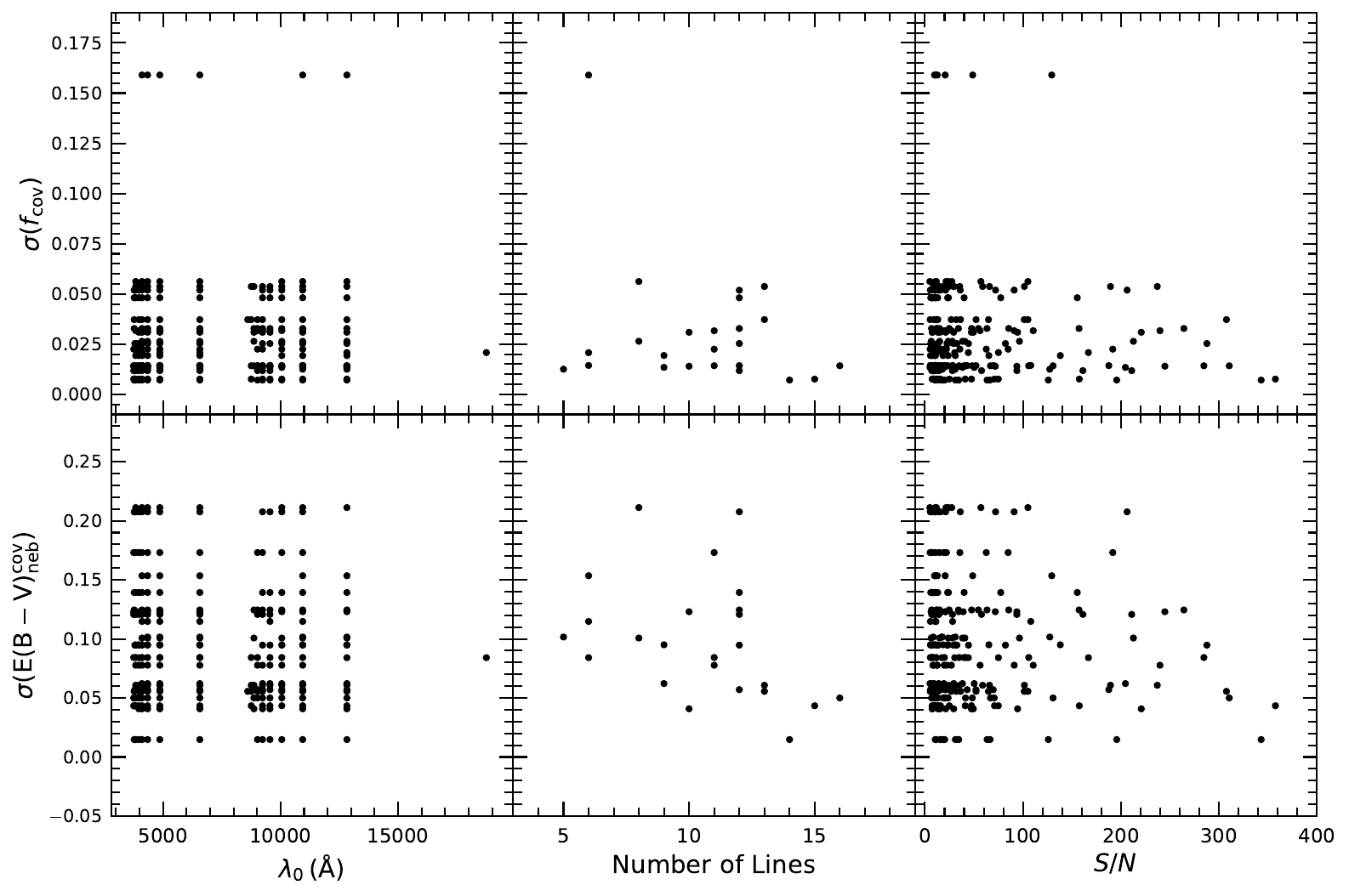}
    \caption{Uncertainties in $\fcov$ (top row) and $\ebmvneb^{\rm
        cov}$ (bottom row) versus line wavelength (left column),
      number of lines (middle column), and line $S/N$ (right column)
      for the 24 galaxies in the sample.}
    \label{fig:subunity_uncertainties}
\end{figure}

Figure~\ref{fig:subunity_parms} shows that, on average, the
uncertainties in both $\ebmvneb^{\rm cov}$ and $\fcov$ increase with
$\fcov$.  Here, we examine whether these larger uncertainties are
driven by a limited baseline in wavelength of the fitted lines, a
lower number of fitted lines, and/or a lower $S/N$ in the lines.
Figure~\ref{fig:subunity_uncertainties} shows the uncertainties in
$\fcov$ and $\ebmvneb^{\rm cov}$ plotted against the wavelength
distribution, number, and $S/N$ of the Balmer and Paschen
recombination emission lines for each of the 24 galaxies in the
sample.  From the top left panel of the figure, it is apparent that
the wavelength coverage of the Balmer and Paschen lines used to fit
$R$ versus $\lambda_0$ for the one galaxy with the largest
$\sigma(\fcov)$ is similar to the wavelength coverage accessible for
galaxies with the lowest $\sigma(\fcov)$.  However, this galaxy
has a lower wavelength density of both higher-order Balmer and
higher-order Paschen lines relative to other galaxies in the
sample.  Thus, while the total number of lines does not appear to be a
strong factor in predicting $\sigma(\fcov)$ (top middle panel), the
wavelength density of lines does appear to play a role.  The top right
panel clearly indicates that the galaxy with the largest
$\sigma(\fcov)$ has fewer higher $S/N$ lines relative to galaxies
with lower $\sigma(\fcov)$.  This galaxy is sufficiently faint
that there are fewer high $S/N$ lines, and hence fewer of the weaker
higher-order Balmer and Paschen lines with $S/N>5$, available for our
analysis.  The same general conclusions hold for galaxies with higher
$\sigma(\ebmvneb^{\rm cov})$ (bottom row of
Figure~\ref{fig:subunity_uncertainties}).

\begin{figure}
  \epsscale{1.2}
  \includegraphics[width=1.0\linewidth]{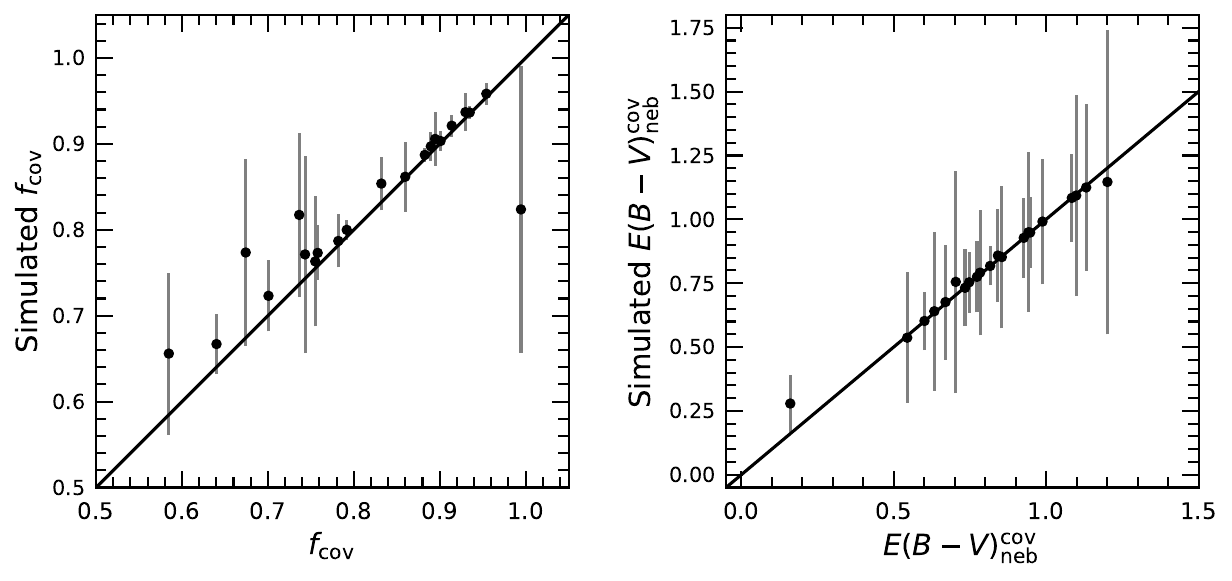}
    \caption{Comparison between $\fcov$ and $\ebmvneb^{\rm cov}$
      deduced from the measured lines and those obtained when
      degrading the $S/N$ of all lines by a factor of 3 and removing
      any lines with $S/N<5$.}
    \label{fig:simsubunity}
\end{figure}

We next consider whether the covering-fraction modeling is biased
towards lower $\fcov$ for galaxies with lower $S/N$ lines and a
sparser wavelength coverage of those lines.  To test for such an
effect, the $S/N$ of all the lines for every galaxy in the sample was
degraded by simply multiplying the flux uncertainties by a factor of
3.  This typically resulted in anywhere from $12\%$ to $50\%$ of the
lines for any given galaxy falling below $S/N=5$.  These lines were
removed (leaving anywhere between 5 and 13 lines\footnote{Three
  galaxies had only 2 to 4 lines remaining after this procedure,
  falling below the minimum of 5 lines required for our analysis, and
  so were removed from this test.}), and the fluxes and flux
uncertainties of the remaining lines with $S/N\ge 5$ were used to
calculate $R$ and $\sigma_{R}$ and then re-determine $\fcov$ and
$\ebmvneb^{\rm cov}$ and their respective uncertainties.
Figure~\ref{fig:simsubunity} compares $\fcov$ and $\ebmvneb^{\rm cov}$
obtained with the original data to the values obtained with the
degraded-$S/N$ lines.  The effect of degrading the lines and removing
any with $S/N<5$ does not introduce a significant systematic offset
compared to the measurement uncertainties for either $\fcov$ or
$\ebmvneb^{\rm cov}$.  If anything, there is a slight tendency for the
covering-fraction model to settle on slightly {\em higher} $\fcov$
relative to the input value when degrading the $S/N$ (and number) of
lines.  Thus, the results of this test suggest that any bias towards
finding a lower $\fcov$ for galaxies with lower $S/N$ and/or fewer
lines is not significant, particularly when compared with the
measurement uncertainties.  In other words, higher $S/N$ spectra of
those galaxies with larger uncertainties in $\fcov$ and $\ebmvneb^{\rm
  cov}$ is unlikely to substantially shift the best-fit $\fcov$ and
$\ebmvneb^{\rm cov}$ to higher and lower values, respectively.  As
discussed in Paper II, galaxies with lower $\fcov$ are
associated with lower-SFR galaxies.

\end{document}